\documentclass[AMA,STIX2COL]{WileyNJD-v2}
\usepackage{moreverb}

\usepackage{makecell}
\usepackage{enumerate}
\usepackage{tabularx}
\usepackage{graphics}
\usepackage{subfig}

\newcommand\BibTeX{{\rmfamily B\kern-.05em \textsc{i\kern-.025em b}\kern-.08em
T\kern-.1667em\lower.7ex\hbox{E}\kern-.125emX}}

\articletype{Article Type}%

\received{<day> <Month>, <year>}
\revised{<day> <Month>, <year>}
\accepted{<day> <Month>, <year>}

\begin{document}

\title{Degree irregularity and rank probability bias in network meta-analysis}

\author[1]{Annabel L. Davies}

\author[1,2]{Tobias Galla}

\authormark{DAVIES and GALLA}

\address[1]{\orgdiv{Theoretical Physics, Department of Physics and Astronomy, School of Natural Sciences}, \orgname{The University of Manchester}, \orgaddress{\state{Manchester}, \country{UK}}}

\address[2]{\orgdiv{Instituto de F\'isica Interdisciplinar y Sistemas Complejos}, \orgname{IFISC (CSIC-UIB), Campus Universitat Illes Balears}, \orgaddress{\state{Palma de Mallorca}, \country{Spain}}}

\corres{Annabel L. Davies, Theoretical Physics, Department of Physics and Astronomy, School of Natural Sciences, The University of Manchester, Manchester M13 9PL, United Kingdom. \email{annabel.davies@postgrad.manchester.ac.uk}}

\jnlcitation{\cname{%
\author{Davies AL} and
\author{Galla T}} (\cyear{2020}), 
\ctitle{Degree irregularity and rank probability bias in network-meta analysis}}

\abstract[Abstract]{Network meta-analysis (NMA) is a statistical technique for the comparison of treatment options. The nodes of the network are the competing treatments and edges represent comparisons of treatments in trials. Outcomes of Bayesian NMA include estimates of treatment effects, and the probabilities that each treatment is ranked best, second best and so on.  How exactly network geometry affects the accuracy and precision of these outcomes is not fully understood. Here we carry out a simulation study and find that disparity in the number of trials involving different treatments leads to a systematic bias in estimated rank probabilities. This bias is associated with an increased variation in the precision of treatment effect estimates. Using ideas from the theory of complex networks, we define a measure of `degree irregularity' to quantify asymmetry in the number of studies involving each treatment. Our simulations indicate that more regular networks have more precise treatment effect estimates and smaller bias of rank probabilities. We also find that degree regularity is a better indicator of NMA quality than both the total number of studies in a network and the disparity in the number of trials per comparison. These results have implications for planning future trials. We demonstrate that choosing trials which reduce the network's irregularity can improve the precision and accuracy of NMA outcomes.
}


\maketitle

\section{Introduction}
\label{intro}

Meta-analysis is an important statistical technique used to combine the results of multiple randomised controlled trials. Often, individual trials have small sample sizes and involve subjects taken from a reduced population. Because of this, it is desirable to systematically integrate results from different trials that address the same clinical question. Over the last four decades meta-analysis has therefore become invaluable for the comparison of treatment options \cite{SALANTI:2012}.

Conventional meta-analysis focuses on pairwise comparisons of treatments. More recently however, network meta-analysis (NMA) has emerged as a technique for making inferences about multiple competing treatments. NMA allows one to combine data from multiple trials even when different trials test different sets of treatment options. The term `network meta-analysis' derives from a graphical representation of the treatments and trials. The nodes of the network graph are the different treatment options and the connecting edges represent comparisons made between the treatments in the trials. NMA combines both direct and indirect evidence for the assessment of treatments. This makes it possible to compare treatments that have not been tested together in any trial \cite{TSD2, DIAS:2018, Lu:Ades:2004, SALANTI:2012}.  

Bayesian NMA in particular has undergone substantial development over recent years. At the same time, it is also recognised that further research is required to fully understand its limitations and to improve the method \cite{Hoaglin:2011, PRISMA:2015}. In this context, simulation studies are frequently used to evaluate the performance of NMA and the factors affecting its accuracy \cite{Pateras:2018}. This approach involves setting up a  model (e.g. a fixed-effect or random-effect model \cite{HARDY:1996, TSD2}) with parameters whose numerical values can be fixed at the beginning. The model is used to produce synthetic (computer-generated) trial data \cite{Morris:2019}. Estimates of the model parameters are then obtained by feeding this synthetic data into the NMA method. The outcome of the NMA can then be compared against the known model parameters. 

A key strength of this approach is the ability to systematically vary the parameters of the model. For example, different relative treatment effects can be explored, or the structure of the network of treatments and trials can be changed. This allows one to systematically investigate the performance of NMA in a range of different scenarios, and to determine the nature of any inaccuracies or biases. To do this, however, Bayesian NMAs must be carried out for many realisations of the synthetic trial data. The overall computational effort can be considerable because the NMA method relies on extensive Markov Chain Monte Carlo sampling  \cite{Geyer:2011, Lunn:2000}.  
 
The primary outcomes of an NMA are estimates of relative treatment effects, and the corresponding credible intervals. Bayesian NMA allows one to rank treatments based on these relative effects, providing a convenient summary for clinical decision making. However, simply ranking treatments as best, second best and so on can be misleading as it does not take into account the level of overlap between credible intervals \cite{Bafeta:2014}. 

As a consequence, a number of other metrics have been developed to compare treatments. One such metric focuses on so-called `rank probabilities'. These quantify the degree of certainty with which each treatment is believed to be the most effective, second most effective, etc., based on the available trial data. Results are often reported in terms of so-called SUCRA values (`surface under the cumulative ranking curve'). These values condense rank probabilities into a numerical summary \cite{Salanti:2011}, and reflect both the magnitude and the uncertainty of treatment effect estimates \cite{PRISMA:2015, Trinquart:2016, Rucker2015}.

Ranking methods have attracted considerable interest in recent years \cite{Trinquart:2016, Veroniki:2016, Veroniki:2018, Daly:2019, Chaimani:2019, Chiocchia:2020}. It is generally recognised that the accuracy of ranking statistics and the treatment effect estimates are likely to be affected by the geometry of the network of treatments and trials \cite{Salanti:2008, Salanti:2008b, Veroniki:2018, Dequen:2014, Hoaglin:2011}. PRISMA guidelines (`Preferred Reporting Items for Systematic Reviews and Meta-Analyses') therefore recommend that authors provide graphical and qualitative descriptions of network geometry \cite{PRISMA:2015}.

A previous simulation study found that the probability of being ranked first is overestimated for the treatment that is tested in the fewest studies in a given network, and underestimated for the treatment included in the most studies \cite{Kibret:2014}. It has been suggested that this is due to differences in the precision of treatment effect estimates \cite{Rucker2015}. Overall it is generally accepted that reporting only the probability of being best can lead to erroneous conclusions \cite{Veroniki:2018, Jansen:2014, Trinquart:2016}. Indeed, the current advice from the PRISMA guidelines is to report the probability that each treatment has each rank \cite{PRISMA:2015}. 

In practice however, the most common ranking statistic in NMAs continues to be the probability that each treatment is ranked best \cite{Petropoulou:2017}. Previous research on the utility of rank probabilities has also focused almost exclusively on the probability of being ranked best. As a result there is very limited evidence on the validity of reporting the full set of rank probabilities or the SUCRA values. Furthermore, due to a lack of appropriate data-generating models and the high computing power required to carry out Bayesian NMAs, simulation studies have been limited to fixed-effects models or networks of two-arm trials only \cite{Kibret:2014, Jonas:2013}. Some progress been made in relating characteristics of network geometry to the outcome of NMA \cite{Salanti:2008b,Salanti:2008,Tonin:2019}. However, it is largely unexplored how exactly these metrics relate to the performance of ranking statistics and treatment effect estimates \cite{Hoaglin:2011}. 

The purpose of our work is to study how the structure of the network affects the probability that each treatment is ranked first, second and so on. In particular we go beyond the probability of being ranked best. We also investigate the mechanisms by which the network affects rank probabilities. Building on recent advances in data-generating methods\cite{Seide:2019}, our simulation studies include  random-effects models and networks of multi-arm trials. In order to characterise network geometry we introduce a measure of asymmetry in the number of studies per treatment which we call `degree irregularity'. The network is said to be regular if all treatments are tested in the same number of studies, and it becomes increasingly more irregular the more this number varies across treatments. Through simulations of multiple network geometries we investigate how this metric affects the precision and accuracy of the treatment effect estimates, and the quality of rank probability estimates and SUCRA values. These results provide a simple method for the identification of additional trials which best complement an existing network of evidence.  

The remainder of this paper is set out as follows: In Section \ref{methods} we present the relevant background information and methods. We begin by outlining the random-effects model for a network of multi-arm trials and the Bayesian approach to NMA. Following this we define the key outcomes of NMA and the relevant ranking statistics. The design of the simulation and networks are described along with details of the data-generating models. We also introduce treatment-specific and network-level quantities that allow us to compare the quality of NMA outcomes within and between networks. Section \ref{results} contains our main results. First, we present within-network and between-network comparisons for networks with equally effective treatments and two-arm trials. We then test how well these results generalise to scenarios in which the true treatment effects vary across treatments, and we study networks involving multi-arm trials. We also compare the results from three different data-generating models. In Section \ref{discuss} we summarise and discuss our main findings. We provide an example that demonstrates how our results can be used to inform the choice of future trials.

\section{Methods}
\label{methods}
\subsection{General setup: network of trials}
  We consider a collection of $N$ treatments, which we label $\alpha =T_1, T_2,\dots, T_N$. The network contains $M$ trials, denoted $i=1,\dots,M$. Each trial compares a subset of treatments, $A_i\subset\{T_1,\dots, T_N\}$; $m_i=|A_i|$ is the number of treatments in trial $i$. We use the notation $t_{i,\ell}$ to label the treatments compared in trial $i$, where $\ell=1,\dots,m_i$. Each $t_{i,\ell}$ is therefore a treatment from the set  $\{T_1,\dots,T_N\}$. 

\begin{figure}
    \centering
    \includegraphics[width=1.0\linewidth]{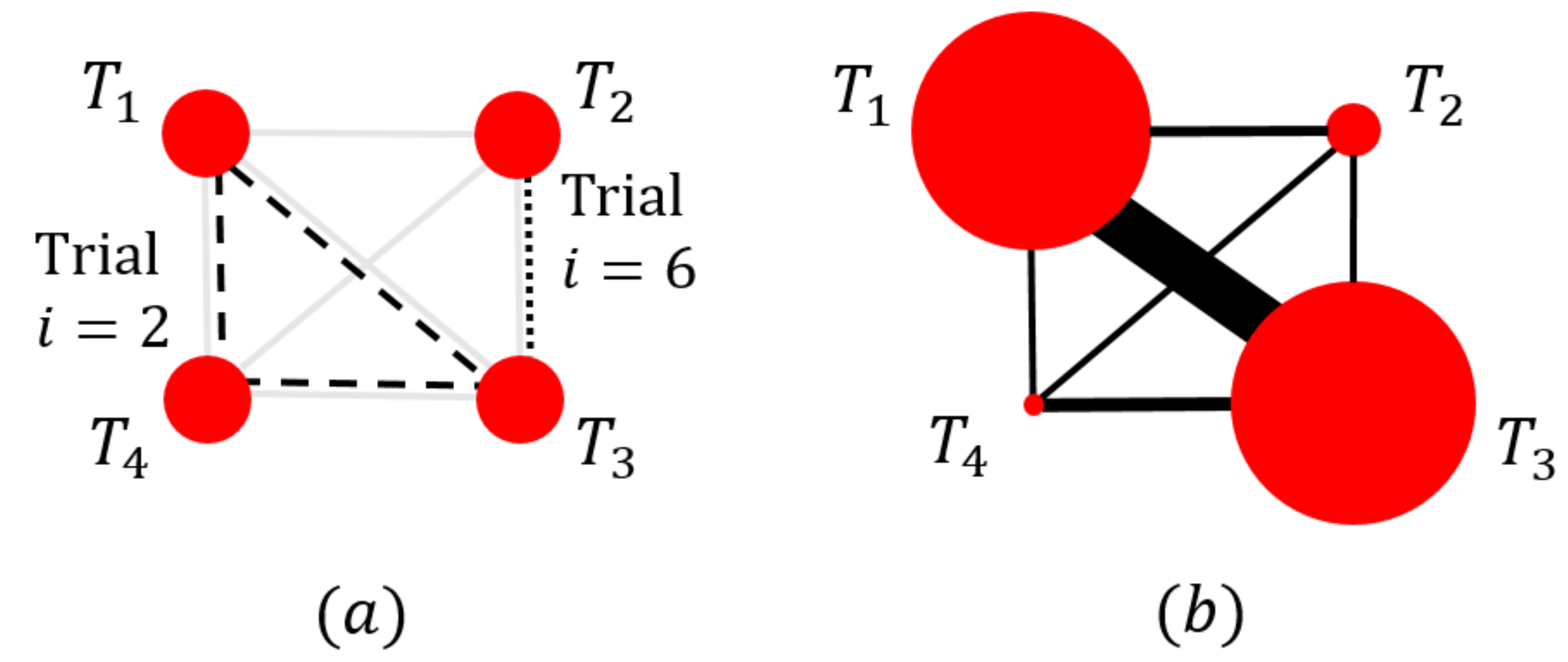}
    \caption{~  Graphical representation of the network of treatments and trials for smoking cessation  \cite{Hasselblad:1998}.The four treatments are: $T_1=$ no contact (control), $T_2=$ self-help, $T_3=$ individual counselling and $T_4=$ group counselling. In panel (a), trials $i=2$ (a 3-arm study) and $i=6$ (a 2-arm study) are highlighted by dashed and dotted lines respectively. The thickness of each egde in panel (b) is proportional to the number of studies that make that comparison, and the diameter each node is proportional to the number of participants who received that treatment. }
    \label{fig:Smoking}
\end{figure}

As an example, consider a network of smoking cessation data reported by Hasselblad \cite{Hasselblad:1998}. Four treatments are compared: $T_1=$ no contact (control), $T_2=$ self-help, $T_3=$ individual counselling and $T_4=$ group counselling. The network is shown in Figure~\ref{fig:Smoking} and consists of 24 trials. Trial $i=2$ in Hasselblad\cite{Hasselblad:1998} compares $m_2=3$ treatments such that $t_{2,1}=T_1$ (no contact), $t_{2,2}=T_3$ (individual counselling) and $t_{2,3}=T_4$ (group counselling). Trial $i=6$, on the other hand, compares $m_{6}=2$ treatments; $t_{6,1}=T_2$ (self-help) and $t_{6,2}=T_3$ (individual counselling). In Figure~\ref{fig:Smoking}(a), these two trials are highlighted by dashed and dotted lines respectively. Showing all $24$ individual trials in this way would result in a rather cumbersome graph. It is therefore common not to indicate trials individually in the network. Instead, two treatments are connected by an edge whenever there is at least one trial involving both treatments. In the example, this means that edges are present between all pairs of treatments. For each pair, the thickness of the link is proportional to the number of trials comparing these two treatments. This turns the graph into a `weighted network' \cite{Newman:2018}. In our illustrations, the diameter of each node is proportional to the number of participants that have received the treatment represented by the node. For the network in Hasselblad\cite{Hasselblad:1998}, this results in Figure \ref{fig:Smoking}(b).

The treatments in trial \(i\) are referred to as the arms of the trial. For a given trial $i$, the treatment in arm $\ell$ is administered to $n_{i,\ell}$ patients. We assume a binary outcome, i.e. the application of the treatment to a particular patient either produces an `event', or it does not. The number of resulting events, $r_{i,\ell}$, is then recorded for each trial and arm.  This means that trial $i$ is defined by the treatments  it compares, $A_i=t_{i,1},\dots, t_{i,m_i}$, and by the number of patients in each arm, $n_{i,\ell}$. The trial reports dichotomous data of the form $(r_{i,1},...,,r_{i,m_i})$. 

The model assumes that the application of the treatment in arm $\ell$ of trial $i$ generates events with probability $p_{i,\ell}$ independently for each of the $n_{i,\ell}$ patients at the end of this trial arm \cite{Hamza:2008}. As a consequence of this setup, each $r_{i,\ell}$ is a binomial random variable,
\begin{equation}
\label{eq:Binomial}
    r_{i,\ell}\sim \text{Bin}(n_{i,\ell},p_{i,\ell}), 
\end{equation}
for $i=1,\dots,M$ and $\ell=1,\dots m_i$.

We use a random-effects model, i.e. $p_{i,\ell}$ may be different from $p_{i',\ell'}$ in different trials ($i\neq i'$), even if $t_{i,\ell}=t_{i',\ell'}$. That is to say, the effectiveness of any fixed treatment $\alpha\in\{T_1,\dots,T_N\}$  may be different in different trials.

Following the generalised linear model framework, the probabilities $p_{i,\ell}$ are described on the logit scale \cite{McCullagh:1989}. We use the notation $\text{logit}(p)=\ln p - \ln (1-p)$ for $0<p<1$. 

Our analysis focuses on relative rather than absolute treatment effects. To this end, we refer to treatment $t_{i,\ell=1}$ as the `baseline' treatment of trial $i$. We write $b_i=\text{logit} ~p_{i,1}$ for the absolute treatment effect of this trial-specific baseline. For $\ell\neq 1$ we then define the relative treatment effect $\delta_{i}(\ell)$, 
\begin{equation}
\label{eq:logit}
\delta_i(\ell)\equiv \text{logit}(p_{i,\ell})-b_{i}.
\end{equation}
 The trial-specific baseline treatment effect, $b_i$, is the log odds of the outcome in arm $\ell=1$ of trial $i$, while the relative treatment effect is the log odds ratio of treatment $\ell\neq 1$ compared to the trial-specific baseline.

\subsection{Random-effects model}
A contrast-based approach to the random-effects model assumes the exchangeability of relative, rather than absolute treatment effects \cite{DerSimonian:1986, Hong:2015, Dias:2016}. This indicates that the relative effect of two treatments $\alpha$ and $\beta$ is drawn from the same distribution for any trial involving these two treatments, no matter what value the index $i$ takes for this trial. The function of the index $i$ is therefore purely to distinguish trials, it does not contain any other a-priori information about the effectiveness of treatments in that trial.

We assume that the relative treatment effects for a given trial $i$ are drawn from a multivariate normal distribution 
\begin{equation}
\label{eq:multi_norm}
  \begin{pmatrix}
    \delta_i(2)\\
    \vdots \\
    \delta_i(m_i)
    \end{pmatrix}\sim \mathcal{N}\left(\begin{pmatrix}
    d_{t_{i,1}t_{i,2}}\\
    \vdots \\
    d_{t_{i,1}t_{i,m_i}}
    \end{pmatrix},\boldsymbol{\Sigma_i}
    \right).
\end{equation}
This means that the relative effect $\delta_i(\ell)$ of the $\ell$-th treatment in trial $i$ (compared to the baseline treatment of trial $i$) is drawn from a Gaussian distribution with mean $d_{t_{i,1},t_{i,\ell}}$. The latter quantity is the mean effect of treatment $t_{i,\ell}$ relative to the baseline treatment $t_{i,1}$ of trial $i$. That is to say, it is the average relative treatment effect one would see in a large sample of trials comparing these two treatments. We assume that these unknown mean relative treatment effects fulfill the consistency relations
\begin{equation}
\label{eq:consistency}
    d_{\alpha\beta}=d_{\alpha\gamma}-d_{\beta\gamma}.
\end{equation}
The $(m_i-1)\times (m_i-1)$ covariance matrix $\boldsymbol{\Sigma_i}$ in Equation~(\ref{eq:multi_norm}) describes the between-trial variance of the relative treatment effects, and their correlations. Following References \cite{Hig:White:1996, Lumley:2002, Lu:Ades:2004}, we will assume that its diagonal elements are all identical. We write $\tau^2$ for their common value. This is the variance of each $\delta_i(\ell)$. We will further assume that the covariance between any two treatment effects is $\tau^2/2$ (these are the off-diagonal elements of $\boldsymbol{\Sigma_i}$). This ensures that the relative effect $\delta_i(\ell)-\delta_i(\ell')$ between \textit{any two} treatments $\ell\neq\ell'$ in trial $i$ has variance $\tau^2$.

The aim of network meta-analysis is to estimate the mean treatment effects $d_{\alpha\beta}$ for all pairs $\alpha\neq\beta$, and the heterogeneity parameter, $\tau$.  Given the consistency assumption (\ref{eq:consistency}), not all $d_{\alpha\beta}$ are independent. As a consequence, we can use treatment $\alpha=T_1$ as the overall global baseline treatment, and it is sufficient to estimate $d_{T_1\alpha}$ for $\alpha=T_2,\dots,T_N$ \cite{Lu:Ades:2006}.

\subsection{Bayesian network meta-analysis}
We write $\boldsymbol{d}=(d_{T_1T_2},d_{T_1T_3},\dots,d_{T_1T_N})$ for the vector of mean treatment effects relative to the global baseline. The vector $\boldsymbol{n}$ contains the numbers of patients in all arms in the network and $\boldsymbol{r}$ contains the trial outcomes. Bayesian NMA aims to construct posterior distributions for the model parameters, $(\boldsymbol{d},\tau)$, conditional on the data, $(\boldsymbol{r},\boldsymbol{n})$. This is achieved using appropriate likelihood functions and prior distributions \cite{Smith:1995, Lu:Ades:2004}. We use non-informative prior distributions for the model parameters. Specifically, we assume independent univariate Gaussian distributions ${\cal N}(0,10^4)$ for each of the parameters $b_i$ and $d_{\alpha\beta}$. The prior for $\tau$ is assumed to be a uniform distribution over the interval from $0$ to $5$ \cite{TSD2, Greco:2016}. As described by Smith et al (1995), the joint likelihood function of all the model parameters, $(\{\delta_i(\ell)\},\{b_i\}, \boldsymbol{d}, \tau)$, can be constructed from the product of so-called conditional `parent-child' probability distributions \cite{Smith:1995}. In the case of the random-effects model for dichotomous data we have described, Eqs.~(\ref{eq:Binomial}) to (\ref{eq:multi_norm}) are used to construct a likelihood which is a product of binomial and multi-variate normal distributions.

For this setup, the posterior distributions of the model parameters can usually not be obtained analytically. We therefore rely on MCMC methods, specifically the Metropolis-in-Gibbs algorithm \cite{Robert:2010, Geyer:2011, Lynch:2007, Lunn:2000}. MCMC approaches work by constructing a Markov chain whose stationary distribution is the probability distribution of interest. In Bayesian NMA this is the posterior distribution of the model parameters, $(\boldsymbol{d}, \tau)$, given the data. 

Following Kibret et al (2014) \cite{Kibret:2014}, we used a burn-in of $5\times 10^3$ and a thinning factor of $10$ in our MCMC simulations. Samples were drawn from the posterior distributions for $2\times 10^4$ iterations after burn-in. 

\subsection{Reporting NMA outcomes}

The primary outcomes from an NMA are the final estimates of the model parameters and their uncertainty (the latter is usually indicated by a 95\% credible interval). In addition, Bayesian NMA allows for the calculation of rank probabilities $\{P_{\alpha}(r)\}$. The quantity $P_\alpha(r)$ is the probability that treatment $\alpha$ is ranked $r$-th. At each MCMC iteration the treatments are ranked from best (rank $r=1$) to worst (rank $r=N$) based on the values of $d_{T_1 \alpha}$ sampled at that iteration. After discarding the burn-in, and carrying out thinning as described, the rank probabilities are estimated from the proportion of times each treatment received each rank.

Treatment effect estimates and ranking probabilities become more difficult to interpret as the number of treatments in the network increases \cite{Veroniki:2018, Trinquart:2016}. In order to simplify this information, Salanti et al (2011) introduced a numerical summary, the so-called `surface under the cumulative ranking' curve (SUCRA) \cite{Salanti:2011}. The value of SUCRA for treatment $\alpha$ is defined as
\begin{equation}
\label{eq:SUCRA}
    \text{SUCRA}_{\alpha} = \frac{1}{N-1}\sum_{r=1}^{N-1}F_{\alpha}(r),
\end{equation}
where $F_{\alpha}(r)$ is the probability that treatment $\alpha$ has rank $r$ or better \cite{Salanti:2011, Rucker2015}, 
\begin{equation}
    F_{\alpha}(r) = \sum_{s=1}^{r}P_{\alpha}(s).
\end{equation}
Using the definition of the expected rank,
\begin{equation}
\langle r \rangle_{\alpha} = \sum_{r=1}^{N}rP_{\alpha}(r), 
\end{equation}
it is straightforward to see that \cite{Rucker2015}
\begin{equation}\label{eq:SUCRA_ER}
    \text{SUCRA}_{\alpha} = \frac{1}{N-1}(N-\langle r \rangle_{\alpha}).
\end{equation}
In this notation SUCRA takes values between zero and one. In practice, SUCRA values are often expressed as a percentage: if a treatment is ranked first with probability one then it will have $\text{SUCRA}=1$ (or 100\%), and if it ranks last with certainty, then it will have a $\text{SUCRA}=0$ (or 0\%) \cite{Salanti:2011}.

\subsection{Network design}

\begin{figure}
    \centering
    \includegraphics[width=1.0\linewidth]{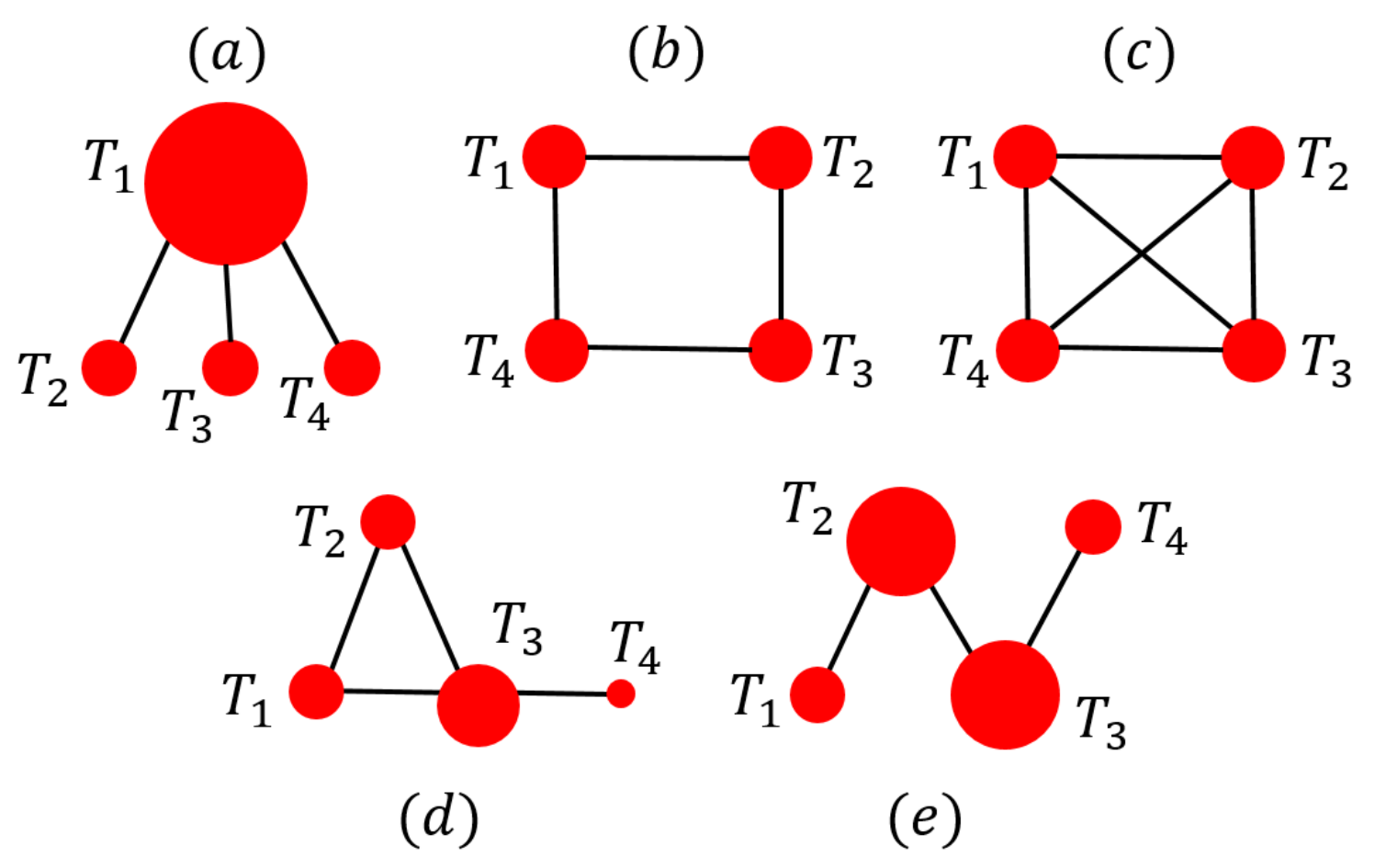}
    \caption{~  Network diagrams of the five network geometries considered in this study: (a) star (b) loop (c) complete loop (d) tadpole (e) ladder. }
    \label{fig:networks}
\end{figure}

Simulations in our work were restricted to networks with $N=4$ treatments. Figure \ref{fig:networks} shows the five network geometries we have used: (a) star, (b) loop, (c) complete loop, (d) tadpole, and (e) ladder. These geometries were chosen as they are commonly observed in real-life network meta-analyses; combinations of these have been previously studied in \cite{Jonas:2013, Kibret:2014, Lu:Ades:2006, Salanti:2008b}.

Within the constraints of these geometries, the number of studies per comparison was varied (i.e., the number of trials involving a particular pair of treatments). To describe the specific geometry of a network we use the vector of the number of studies per comparison, $\boldsymbol{K} = (K_{T_1T_2}, K_{T_1T_3}, K_{T_1T_4}, K_{T_2T_3}, K_{T_2T_4}, K_{T_3T_4})$, where $K_{\alpha\beta}=K_{\beta\alpha}$ is the number of studies that compare treatments $\alpha$ and $\beta$.  The entries of $\boldsymbol{K}$ define the strengths (or weights) of the edges in the network of treatments.

We note, however, that the full setup of the treatment--trial network is not fully specified by $\boldsymbol{K}$ alone. This is because the same number of comparisons per pair of treatments can be achieved by different combinations of two-arm and multi-arm trials.

From $\boldsymbol{K}$ we can obtain the number of studies involving treatment $\alpha$,  
\begin{equation}
  k_{\alpha} = \sum_{\beta \neq \alpha} K_{\alpha \beta}.
\end{equation}
In  the theory of networks this quantity is referred to as the `weighted degree' of node $\alpha$ \cite{Newman:2018}. We will occasionally use slightly more casual language, and refer to $k_\alpha$ as the `number of studies per treatment'. We also define the average number of studies that a treatment is involved in (the `mean degree'),
\begin{equation}
    \hat{k} = \frac{1}{N}\sum_{\alpha}k_{\alpha}.
\end{equation}
 
In network theory, a graph is said to be `regular' if all nodes have the same degree \cite{Newman:2018}. With this in mind, we introduce a measure of `degree irregularity' of the network, 
\begin{equation}
\label{eq:h^2}
h^2 = \frac{1}{N}\sum_{\alpha} (k_{\alpha} - \hat{k})^2.
\end{equation}
This quantifies the variation in the number of studies per treatment. In particular, $h^2=0$ when all nodes are involved in the same number of trials ($k_\alpha=\hat k$ for all $\alpha$). When we make comparisons between networks we use the normalised network irregularity, $h^2/\hat{k}^2$. 

We note that there is a direct mapping between $h^2/\hat{k}^2$ and the so-called `probability of inter-specific encounter index' (PIE). This index is a measure of ecological diversity, and was introduced to network meta-analysis by Salanti et al (2008) \cite{Salanti:2008, Salanti:2008b}. Further details can be found in Section S1 in the Supplementary Material \cite{Davies:Supp}. 
    
Some of the key quantities we use in our analysis are summarised in Table~\ref{Tab:definitions}.  
    
\begin{table}
\caption{~  Summary of key quantities used in our analysis. }
\centering
\begin{tabularx}{\linewidth}{lX}
\toprule
Variable & Definition \\
\midrule
$N$ & Total number of treatments in the network \\
$M$ & Total number of studies in the network \\
$d_{\alpha\beta}$ & True mean relative treatment effect between treatments $\alpha$ and $\beta$ \\
$\tau$ & Heterogeneity parameter  \\
$k_{\alpha}$ & Number of studies involving treatment $\alpha$ \\
$\hat k$ & Mean degree of the network (mean number of studies a treatment is involved in)\\
$h^2/\hat{k}^2$ & Normalised degree irregularity \\
SD$(d)_{\alpha}$ & Treatment-specific standard deviation of the treatment effect estimate \\
$\langle \Delta P_{\alpha}(r)\rangle$ & Bias of the rank probability estimate, averaged over realisations of synthetic trial data \\
$\overline{\text{SD}}$ & Total standard deviation of treatment effect estimates in the network\\
$\overline{|\Delta P|}$ &  Total rank probability bias in the network \\
$\overline{|\Delta \text{SUCRA}|} $ &  Total SUCRA bias in the network \\
\bottomrule
\end{tabularx}
\label{Tab:definitions}
\end{table}
\subsection{Simulation method}
In our simulations we generate dichotomous trial data for a specified network geometry and for known model parameters $(\boldsymbol{d},\tau)$. An NMA is performed for multiple independent realisations of simulated data, and the resulting estimates of the model parameters are recorded for each realisation. More specifically, we used the following numerical protocol:

\begin{enumerate}[(1)]
\item Define the fixed parameters of the network such as the total number of studies, $M$, the vector of number of studies per comparison, $\boldsymbol{K}$, the number of participants in each arm, $n_{i, \ell}$, and the true model parameter values $(\boldsymbol{d},\tau)$. 
\item Generate and analyse independent realisations $\nu=1,2,...,\Omega$ of synthetic trial outcomes. Specifically, for each $\nu$:
\begin{enumerate}[(a)]
    \item For all trials $i$, randomly sample the $\{\delta_i(\ell)\}$, $\ell=2,\dots, m_i$, from the multivariate normal distribution in Equation~(\ref{eq:multi_norm}).
    \item Using the $\{\delta_i(\ell)\}$ and one of the three data-generating models (see Section \ref{DGM}), construct the probabilities $p_{i,\ell}$, $\ell=1,\dots,m_i$, for all trials $i$ in the network.
    \item For each trial arm, generate random event data, $r_{i,\ell}$, from the binomial distribution in Equation~(\ref{eq:Binomial}).
    \item Use the vector of events, $\boldsymbol{r}$, and vector of participants, $\boldsymbol{n}$, to carry out a Bayesian NMA.
    \item Determine the treatment effects with respect to the baseline $T_1$ and use the consistency relation in Equation (\ref{eq:consistency}) to output the estimated model parameters, $\Tilde{d}_{\alpha\beta}^{(\nu)}$, for all $\alpha,\beta \in\{T_1,\dots, T_N\}$.  Also output the estimated heterogeneity parameter, $\Tilde{\tau}^{(\nu)}$, and the bias of rank probabilities,
    \begin{equation}
    \label{eq:biasP}
        \Delta P_{\alpha}^{(\nu)}(r) = \Tilde{P}_{\alpha}^{(\nu)}(r)-P_{\alpha}(r).
    \end{equation}
  In this equation $\Tilde{P}_{\alpha}^{(\nu)}(r)$ is the probability that treatment $\alpha$ has rank $r$ in the NMA of realisation $\nu$. 
\end{enumerate}
\item Calculate the mean and standard deviations of the estimated model parameters over realisations, for example
\begin{align}
    \langle \Tilde{d}_{\alpha\beta} \rangle &= \frac{1}{\Omega}\sum_{\nu=1}^{\Omega} \Tilde{d}_{\alpha\beta}^{(\nu)}, \\
    \text{SD}( \Tilde{d}_{\alpha\beta} ) &= \sqrt{\frac{1}{\Omega-1} \sum_{\nu=1}^{\Omega} (\Tilde{d}_{\alpha\beta}^{(\nu)} - \langle \Tilde{d}_{\alpha\beta} \rangle )^2 },
\end{align}
with similar definitions for $\langle \tilde{\tau} \rangle $, $\text{SD}(\Tilde{\tau})$, $\langle \Delta P_{\alpha}(r) \rangle $ and $\text{SD}(\Delta P_{\alpha}(r))$.
\end{enumerate}

\subsection{Data generation for simulation studies}
\label{DGM}
The relative treatment effects $\delta_i(\ell)$ ($\ell=2,\dots, m_i$) in any one trial $i$ do not uniquely define the absolute treatment effects $p_{i,\ell}$. This is because the treatment effect of the trial-specific baseline, $p_{i,1}$, is not determined by the $\{\delta_i(\ell)\}$. Equation~(\ref{eq:logit}) can be re-arranged to give
\begin{equation}\label{eq:p_i_ell}
    p_{i,\ell}=p_{i,\ell}[p_{i,1},\delta_i(\ell)] = \frac{p_{i,1}\exp(\delta_i(\ell))}{1-p_{i,1}\left(\exp(\delta_i(\ell))-1\right)},
\end{equation}
so that $p_{i,1}$ together with the $\{\delta_i(\ell)\}$ ($\ell=2,\dots, m_i$) specifies all absolute treatment effects in trial $i$.

To fully define step (2)(b) in the above algorithm it is therefore sufficient to specify the construction of $p_{i,1}$. In the context of the random-effects model and to allow for the inclusion of multi-arm trials, we use three data-generating models (DGM) based on those presented by Seide et al (2019) \cite{Seide:2019}. 

The first DGM, which we will call `Euclidean', chooses the treatment effect for the baseline treatment to be the value that minimises the Euclidean distance of the vector $(p_{i,1},\dots,p_{i,m_i})$ from the vector $(1/2, \dots, 1/2)$, i.e.,
\begin{multline}
    p_{i,1} = \min\limits_{q} \Biggl[\left(q-\frac{1}{2}\right)^2+ \sum_{\ell=2}^{m_i}\bigg(p_{i,\ell}\left[q,\delta_i(\ell)\right]-\frac{1}{2}\bigg)^2 \Biggr],
\end{multline}
where $p_{i,\ell}[\cdot,\cdot]$ is the expression given in Equation~(\ref{eq:p_i_ell}). This is referred to as `DGM ``Fixed'' Modified' in Seide et al (2019)\cite{Seide:2019}.  The other two methods are variations of the DGM ``Fixed" in Seide et al (2019)\cite{Seide:2019} which we will refer to as `Uniform' and `Normal' respectively. The former samples $p_{i,1}$ from a uniform distribution between zero and one, whilst the latter samples it from a normal distribution $\mathcal{N}(0.5,0.2)$, truncated at zero at the lower end, and at one at the upper end. To ensure our results were not due to the data-generating model, all simulations were performed using each method and the results compared. 

\subsection{Quantities indicating and characterising NMA quality}
In this section we introduce quantities that measure the quality of the parameter estimates resulting from the NMA. We begin with treatment specific values. These are used to compare how well an NMA estimates the effectiveness of each treatment within a given network.

For each pair of treatments $\alpha\neq\beta$ we define the mean bias of the relative treatment effect,
\begin{equation}
    \langle \Delta  d_{\alpha\beta} \rangle = \langle \Tilde{d}_{\alpha\beta} \rangle - d_{\alpha\beta}. 
\end{equation}
For each fixed treatment $\alpha$ we can also define the treatment-specific mean bias 
\begin{align}
 \langle \Delta d \rangle_{\alpha}
 =\frac{1}{N-1}\sum_{\beta \neq \alpha} \langle \Delta d_{\alpha\beta} \rangle.
 \end{align}
Similarly, the treatment-specific standard deviation of the treatment effect is defined as 
 \begin{equation}
 \text{SD}(d)_{\alpha}=\frac{1}{N-1}\sum_{\beta \neq \alpha} \text{SD}(\tilde{d}_{\alpha\beta}).
 \end{equation}
Bias of SUCRA$_{\alpha}$ values and bias of probability ranks, $\langle \Delta P_{\alpha}(r) \rangle$, are treatment specific quantities by construction. The former can be written in terms of the latter, 
 \begin{align}
    \langle \Delta \text{SUCRA}_{\alpha} \rangle = -\frac{\sum_{r} r\langle \Delta P_{\alpha}(r) \rangle}{N-1}.
 \end{align}

Next we define network-level indicators allowing comparisons of the quality of NMA outcomes between networks. We introduce the total magnitude of the bias of rank probability,  
\begin{equation}
    \overline{|\Delta P|} = \sum_{\alpha}\sum_{r} | \langle \Delta P_{\alpha r} \rangle |,
\end{equation}
and the total magnitude of the bias of SUCRA,
\begin{equation}
    \overline{|\Delta \text{SUCRA}|} = \sum_{\alpha} |\langle \Delta \text{SUCRA}_{\alpha} \rangle |.
\end{equation}
To be able to compare numerical values for these two quantities with each other, we express these indicators as proportions of the maximum values they can take, see Section S2 in the Supplementary Material \cite{Davies:Supp} for further details.

Finally, we introduce the total standard deviation and total bias of treatment effects,
\begin{align}
    \overline{\text{SD}} &= \sum_{\alpha} \text{SD}(d)_{\alpha}, \\
    \overline{|\Delta d|} &= \sum_{\alpha} |\langle \Delta d \rangle_{\alpha}|.
\end{align}
Refer to  Table~\ref{Tab:definitions} for a summary of some of these quantities.

\section{Results}
\label{results}

The set of simulated networks was chosen to cover a range of values for the degree irregularity $h^2/\hat{k}^2$. It includes all network geometries in Figure \ref{fig:networks}, with  varying values of $\boldsymbol{K}=(K_{T_1T_2},K_{T_1T_3},K_{T_1T_4},K_{T_2T_3},K_{T_2T_4},K_{T_3T_4})$. We used $\tau=0.1$ throughout, as well as an equal number of participants per arm, $n_{i,\ell}=25$, and $\Omega=10^3$ independent realisations of synthetic trial outcomes for any fixed set of model parameters $(\boldsymbol{d},\tau)$. Error bars in our figures are typically smaller than the size of the markers.

In Sections \ref{sec:within} and \ref{sec:between} we first focus on networks with equally effective treatments, $\boldsymbol{d}=(d_{T_1T_2},d_{T_1T_3},d_{T_1T_4})=(0,0,0)$. Any systematic effect observed in the outcome of NMA is therefore a result of the structure of the network only. Networks with treatments of varying effectiveness are discussed in Section~\ref{sec:varying}.

\subsection{Comparisons within networks}\label{sec:within}

\begin{figure}
    \centering
    \includegraphics[width=1.0\linewidth]{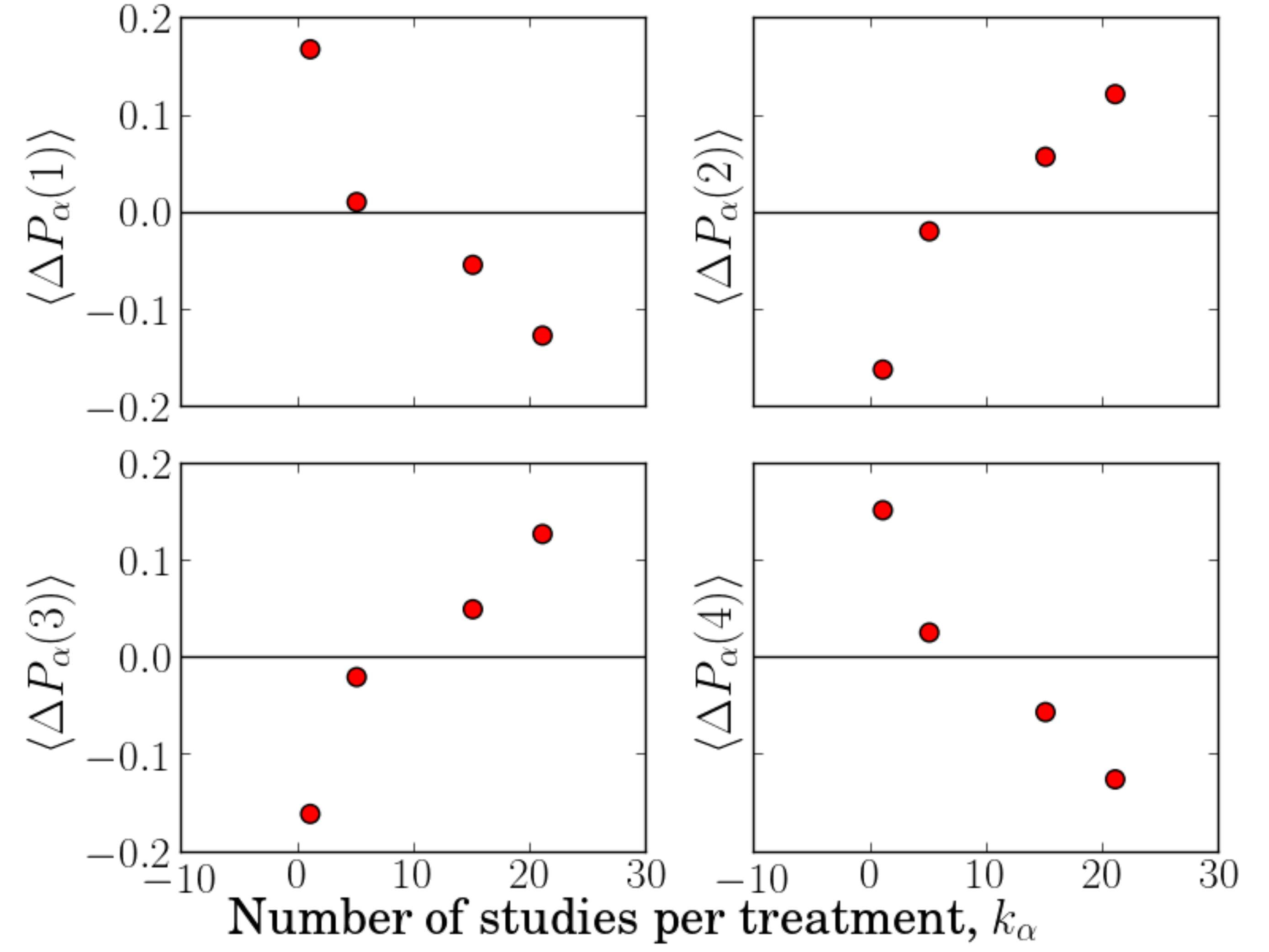}
    \caption{~  The effect of the number of studies per treatment on the bias on rank probabilities, $\Delta P_{\alpha}(r)$, for $r=1,2,3,4$. These plots are for a star network with $\boldsymbol{K}=(1,5,15,0,0,0)$. }
    \label{fig:RankProb_multi}
\end{figure}

In Figure \ref{fig:RankProb_multi} we plot bias of rank probability against number of studies per treatment, $k_\alpha$, for a star network with $\boldsymbol{K}=(1,5,15,0,0,0)$. Similar plots for other network geometries can be found in the Supplementary Material \cite{Davies:Supp} (Figures S1 to S16). This data consistently shows that the probability to be ranked best or worst, $P_{\alpha}(1)$ and $P_{\alpha}(4)$ respectively, is overestimated for the treatment included in the fewest studies (lowest degree $k_\alpha$). The probabilities $P_{\alpha}(2)$ and $P_{\alpha}(3)$ are underestimated. The reverse is found for the treatment included in the most studies. 

\begin{figure}
   \centering
    \includegraphics[width=1.0\linewidth]{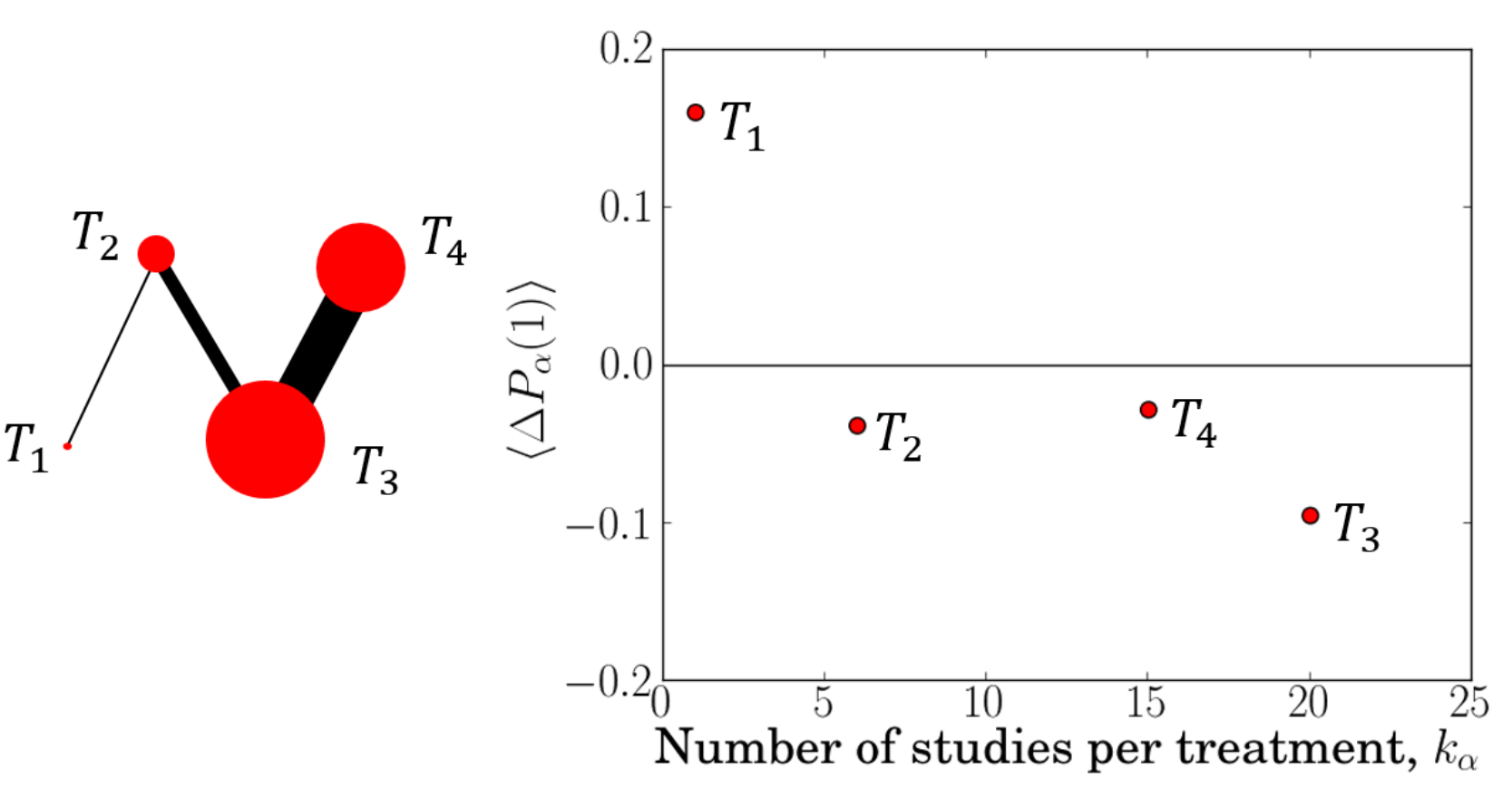}
    \caption{~  Ladder network with $\boldsymbol{K}=(1,0,0,5,0,15)$. An example demonstrating the effect of node position on rank probability bias.}
    \label{fig:Centrality}
\end{figure}

The bias of rank probability for the treatments with the most and fewest studies appears to be common in all networks. We find that the bias of the remaining two treatments can be affected by the position of their respective nodes in the network. Figure \ref{fig:Centrality} shows the bias of $P_{\alpha}(1)$  for a ladder network with $\boldsymbol{K}=(1,0,0,5,0,15)$. In this example treatment $T_2$ is included in fewer studies ($k_{T_2} = 6$) than treatment $T_4$ ($k_{T_4}=15$) but has more direct comparisons (it is directly compared to $T_1$ and $T_3$ whereas treatment $T_4$ is only directly compared to $T_3$). The bias on $P_{\alpha}(1)$ for treatment $\alpha=T_2$ is found to be more negative than that of treatment $\alpha=T_4$. 

\begin{figure}
    \centering
    \includegraphics[width=0.94\linewidth]{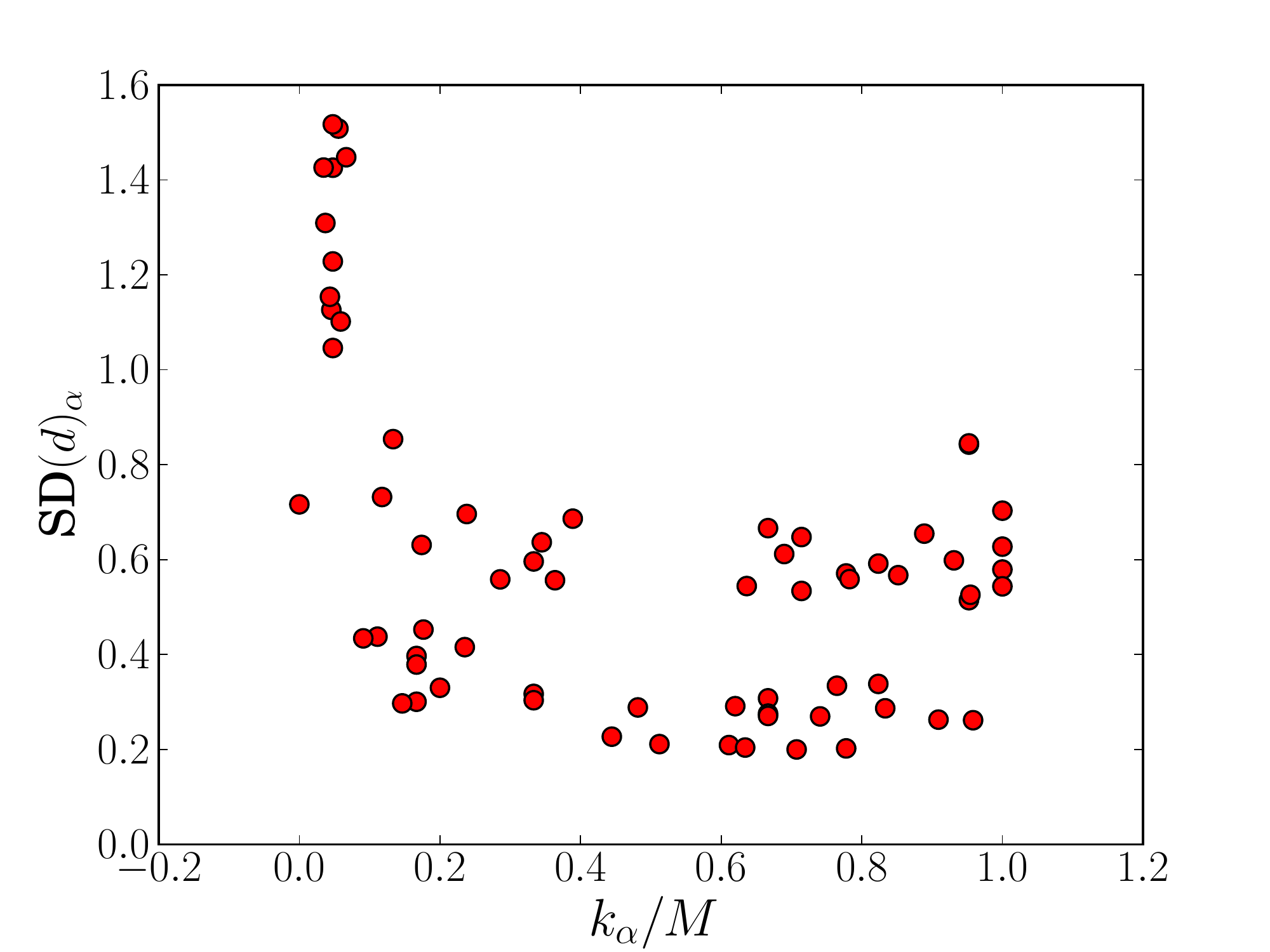}
    \caption{~  Standard deviation of treatment effect estimates. On the horizontal axis, we use the normalised number of studies, $k_\alpha/M$, to capture how well connected a treatment is in the network. Each network contributes four data points, one for each treatment. The figure includes the data from all irregular networks we simulated.}
    \label{fig:SD}
\end{figure}

We conclude that disparity in the number of studies per treatment generates a trend in bias of rank probabilities. It is natural to ask if a similar trend is found for bias of treatment effect estimates. This appears not to be the case (see Figures S1 to S16 in the Supplementary Material \cite{Davies:Supp}). 

Instead, the trend in $\Delta P_{\alpha}(r)$ is associated with a systematic pattern in the standard deviation of treatment effect estimates, see Figure~\ref{fig:SD}.  We find that $\mbox{SD}(d)_{\alpha}$ tends to decrease with the number of studies treatment $\alpha$ is involved in. The standard deviation, $\mbox{SD}(d)_{\alpha}$, is particularly high for treatments with $k_{\alpha}/M \lesssim 0.1$ and appears to flatten out for those included in a larger proportion of studies. We note, however, that Figure \ref{fig:SD} includes data from multiple networks. On inspection of individual networks, we find a slight but consistent decrease in SD($d)_{\alpha}$ as the number of studies of treatment $\alpha$ increases (see Figures S1 to S16 in the Supplementary Material \cite{Davies:Supp}).  This data suggests that bias in the rank probabilities may originate from a variation in the uncertainties of the different treatment effect estimates. A possible mechanism for this is discussed in Section \ref{discuss}.

\subsection{Comparisons between networks}\label{sec:between}
\begin{figure}
  \centering
  \includegraphics[width=0.94\linewidth]{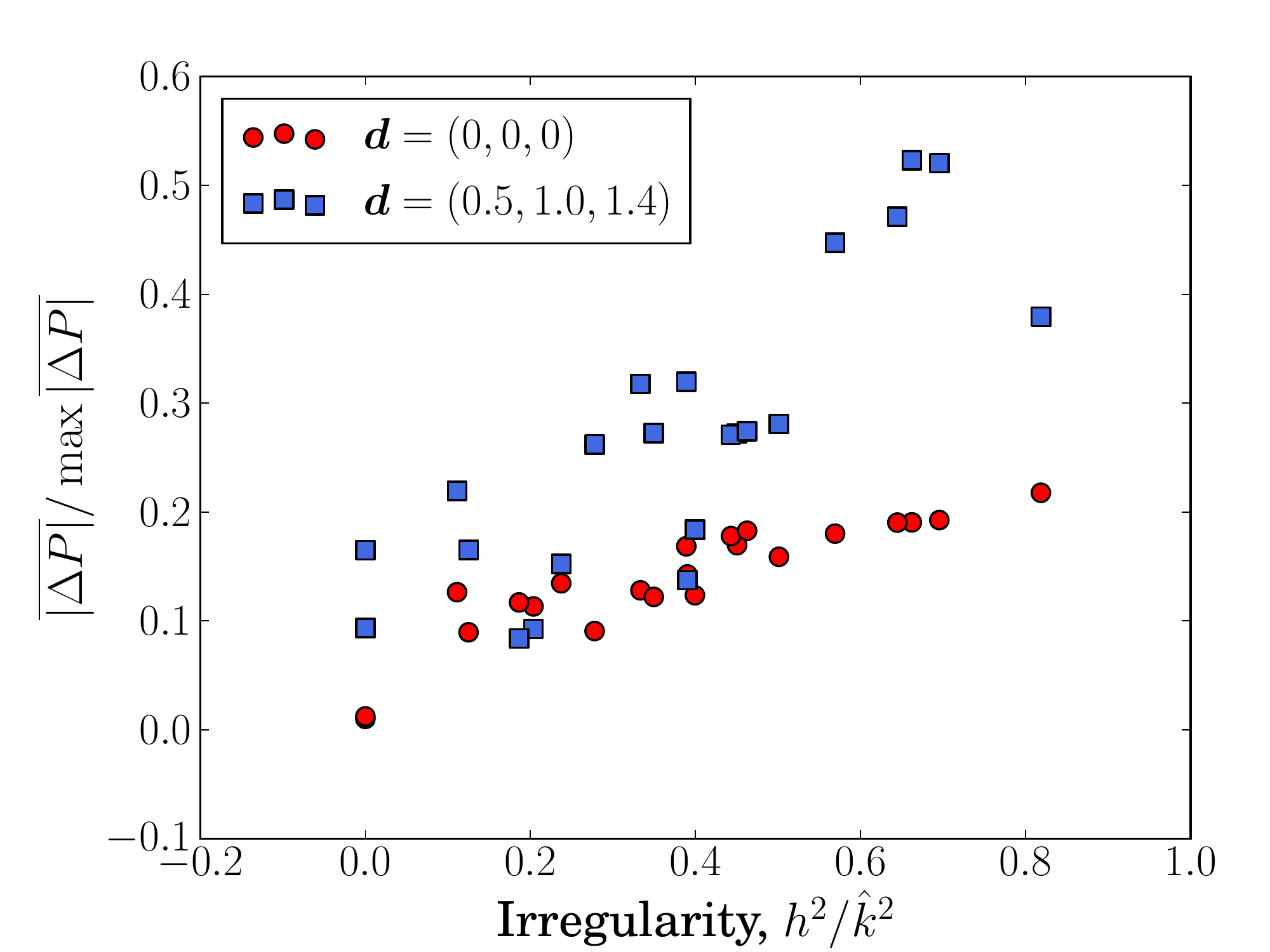}
  \caption{~  The effect of degree irregularity on a network's total rank probability bias for networks with equally effective treatments and non-equally effective treatments.}
  \label{fig:h_probrank_compare}
  \end{figure}

So far we have mostly compared the outcome of NMA for different treatments within a given network. In this section we make comparisons between different networks. One main observation is a positive association between the degree irregularity of a network, $h^2/\hat{k}^2$, and the total bias on rank probabilities, $\overline{|\Delta P|}$. The data shown as red circles in Figure~\ref{fig:h_probrank_compare} demonstrate this for networks with equally effective treatments. 

While we find no relationship between irregularity and total bias of relative treatment effects, $\overline{|\Delta d|}$, (see Figure S21 in the Supplementary Material \cite{Davies:Supp}) Figure~\ref{fig:h_SD_compare} shows that the total standard deviation of the estimates of relative treatment effects, $\overline{\text{SD}}$, increases with $h^2/\hat{k}^2$. Therefore networks with a more homogeneous distribution of studies lead not only to lower bias of rank probabilities, but also to more precise estimates of relative treatment effects. This is also a possible explanation for the vertical spread in Figure ~\ref{fig:SD}. Different data points for a given value of  $k_\alpha/M$ can be from networks with varying degrees of irregularity and hence they result in different outcomes for $\text{SD}(d)_\alpha$

We find that the total bias in rank probability estimates, $\overline{\Delta P}$, and the total standard deviation of treatment effect estimates, $\overline{\text{SD}}$, are not systematically affected by the total number of studies in the network (Figures S22 and S23 in the Supplementary Material \cite{Davies:Supp}). This has implications for the planning of future studies to be added to an existing network. Naively, one may assume that adding any study to an existing network will improve the quality of results because the amount of evidence is increased. However our results suggest that, in terms of bias on rank probabilities and the precision of treatment effect estimates, this is only true if the addition of the study reduces the degree irregularity, $h^2/\hat{k}^2$, of the network.

  \begin{figure}
   \centering
  \includegraphics[width=0.94\linewidth]{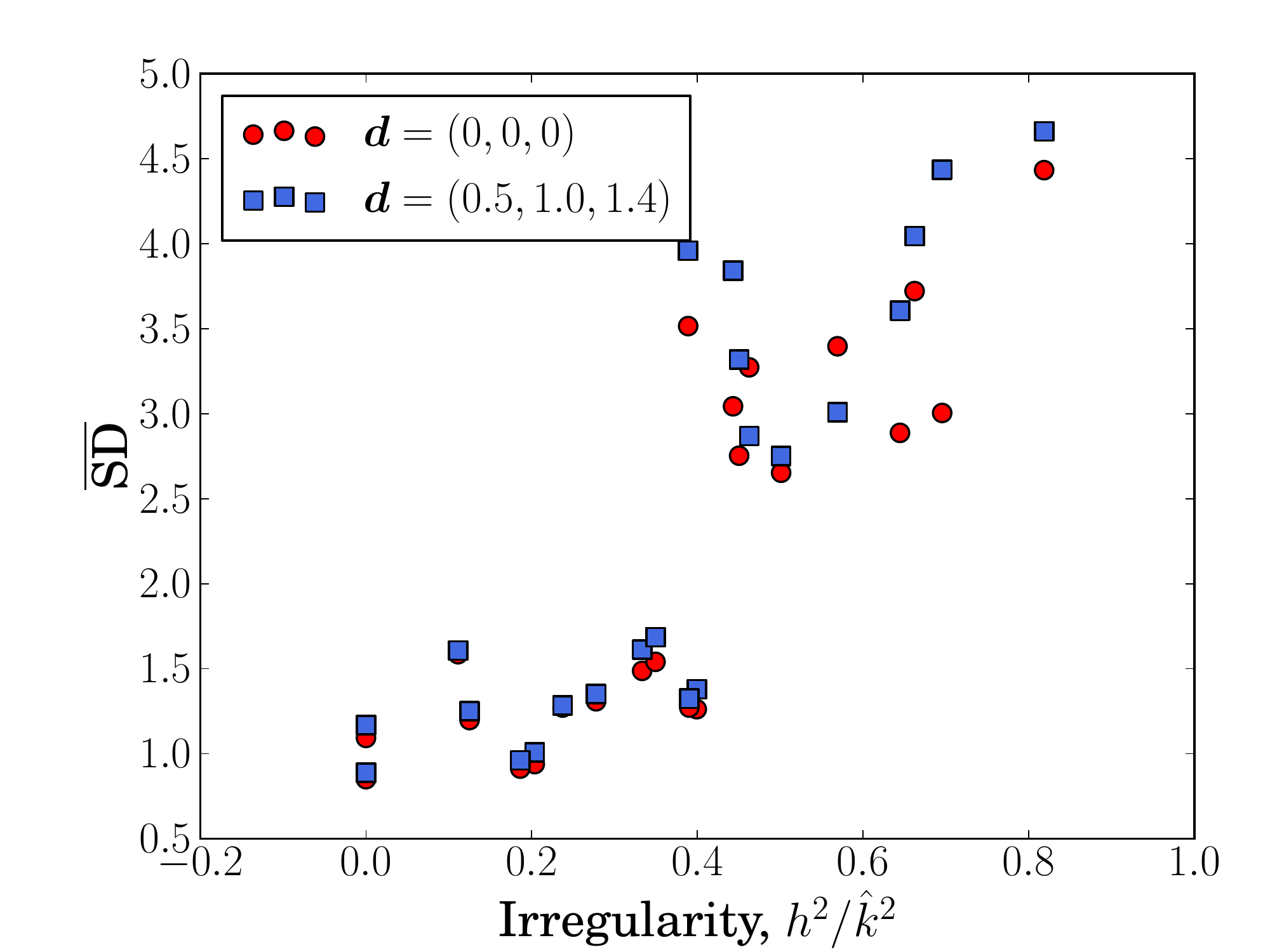}
  \caption{~  The effect of degree irregularity on a network's total standard deviation, $\overline{\text{SD}}$, for networks with equally effective treatments and non-equally effective treatments.}
  \label{fig:h_SD_compare}
  \end{figure}
  
  \begin{figure}
   \centering
  \includegraphics[width=0.94\linewidth]{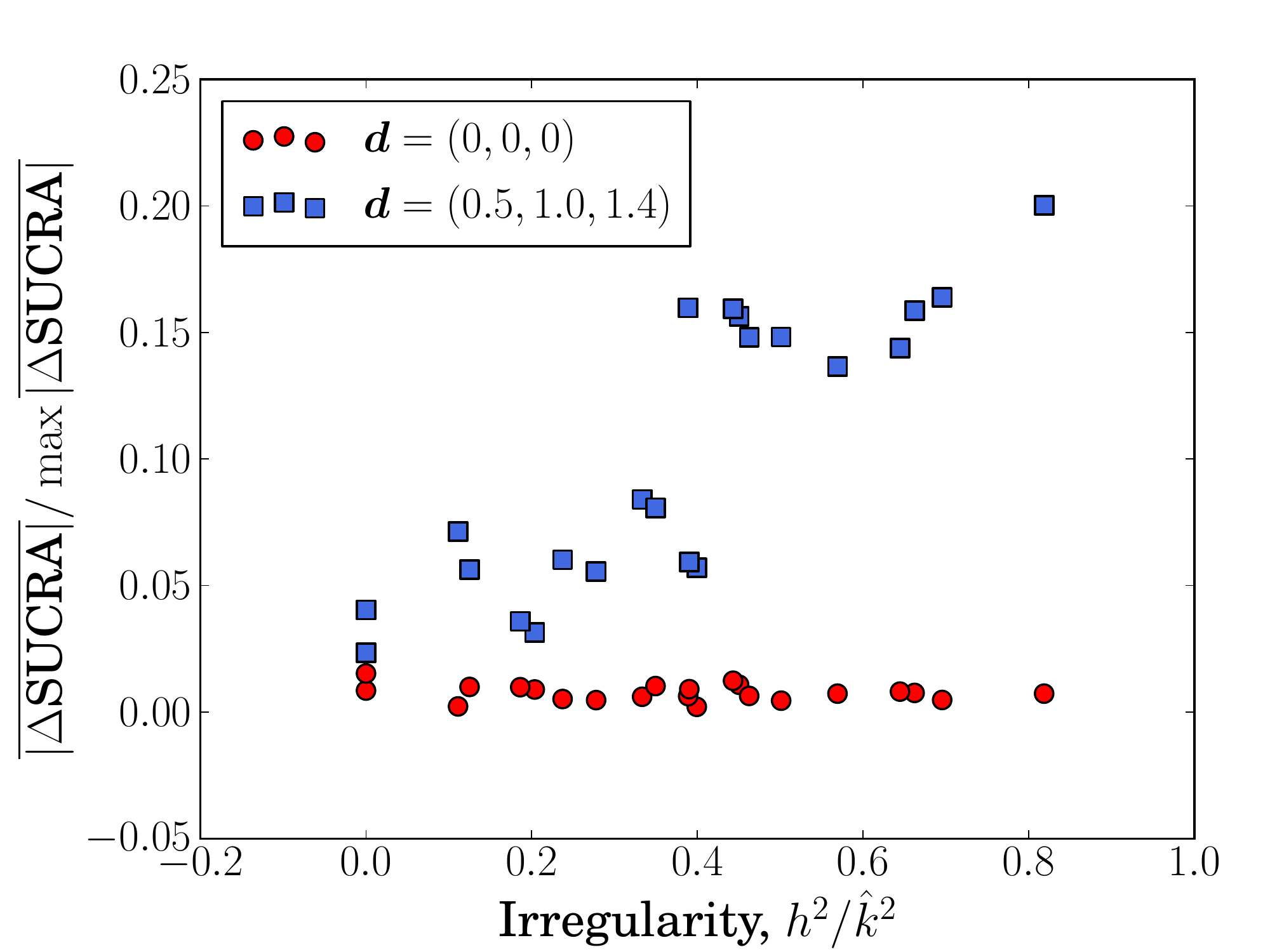}
  \caption{~  The effect of degree irregularity on a network's total SUCRA bias for networks with equally effective treatments and non-equally effective treatments.}
  \label{fig:h_SUCRA_compare}
\end{figure}

The data shown as red circles in Figure~\ref{fig:h_SUCRA_compare} demonstrate that, for networks with equally effective treatments, network irregularity has no effect on the total bias of SUCRA values across the network. Comparing the data in Figures \ref{fig:h_probrank_compare} and \ref{fig:h_SUCRA_compare} we find that the bias of SUCRA is approximately ten times smaller than that of the rank probabilities. This is consistent with the data in Figure~\ref{fig:RankProb_multi} which shows that the biases of $P_{\alpha}(2)$ and $P_{\alpha}(3)$ are almost the exact negative of the biases of $P_{\alpha}(1)$ and $P_{\alpha}(4)$. These biases cancel in calculation of SUCRA in Equation (\ref{eq:SUCRA}). The same reasoning also explains why, when making within-network comparisons, the number of studies per treatment has no effect on the bias of SUCRA$_{\alpha}$ (see Figure S17 in the Supplementary Material \cite{Davies:Supp}).

\subsection{Treatments of varying effectiveness}\label{sec:varying}
The data presented so far is for networks with equally effective treatments, $\boldsymbol{d}=(0,0,0)$. In order to test the robustness of our findings, we now focus on a case in which the four treatments have different effectiveness. Specifically, we choose $\boldsymbol{d}=(0.5,1.0,1.4)$, and study the same network geometries as before. Treatment $T_1$ is now the most effective, followed by $T_2$, then $T_3$ and treatment $T_4$ is the least effective. Therefore the `true' rank probabilities are $P_{T_1}(1)=P_{T_2}(2)=P_{T_3}(3)=P_{T_4}(4)=1.0$ and all other $P_{\alpha}(r)$ are  zero.

As shown in Figure \ref{fig:h_SD_compare}, the relationship between $\overline{\text{SD}}$ and degree irregularity is the same as in the case of equally effective treatments ($\boldsymbol{d}=(0,0,0)$); the numerical values of $\overline{\text{SD}}$ are also found to be largely similar, there is no systematic increase or reduction in the standard deviation.

The qualitative effect of degree irregularity, $h^2/\hat{k}^2$, on the total magnitude of the bias of rank probabilities, $\overline{|\Delta P|}$, is similar compared to the case of equally effective treatments (Figure \ref{fig:h_probrank_compare}). For $h^2/\hat{k}^2 \gtrsim 0.2$ we find that the bias is larger for treatments with varying effectiveness than for equally effective treatments.

In our analysis of the case $\boldsymbol{d}=(0.5,1.0,1.4)$ we find that the standard deviations, SD$(d_{\alpha\beta})$, range from approximately $0.1$ to $1.0$. This means that there is significant overlap in the distributions of the estimated treatment effects. As a consequence of this, the treatments appear to be similarly effective on average. For treatments with varying effectiveness, the true rank probabilities take values of either $0$ or $1$, whereas for networks with equally effective treatments, all $P_{\alpha}(r)$ are equal to $0.25$. It is therefore natural that the bias of rank probabilities is greater for networks of treatments with different effectiveness, at least when the magnitude of SD$(d_{\alpha\beta})$ is of the same order or larger than the disparity in true treatment effects.

 Figure \ref{fig:h_SUCRA_compare} shows that, in contrast to the results for networks with equally effective treatments, $\overline{|\Delta \text{SUCRA}|}$ increases with $h^2/\hat{k}^2$ for $\boldsymbol{d}=(0.5,1.0,1.4)$. On inspection of the biases of rank probabilities within a given network (Figure S20 in the Supplement \cite{Davies:Supp}) we find that the relationship between rank probability bias and the number of studies per treatment is affected by the quality of the treatments that have been compared. Unlike in Figure \ref{fig:RankProb_multi} (where $\boldsymbol{d}=(0,0,0)$), the biases on $P_{\alpha}(2)$ and $P_{\alpha}(3)$ are not equal to $-P_{\alpha}(1)$ and $-P_{\alpha}(4)$ so there is no net cancellation of biases in the calculation of SUCRA$_{\alpha}$. The total bias on rank probabilities increases for more irregular networks (higher values of $h^2/\hat{k}^2$), and as a consequence the bias on SUCRA also increases with the irregularity of the graph.

The data in Figures \ref{fig:h_probrank_compare} to \ref{fig:h_SUCRA_compare} indicate that reducing the network's irregularity improves the precision of treatment effect estimates and reduces bias on ranking statistics in the case of treatments with varying degrees of effectiveness. Our conclusions regarding the use of network heterogeneity for the planning of future studies are therefore also valid in this more realistic scenario.

\subsection{Multi-arm trials}
\label{multi-arm}
The results presented so far are for networks made up exclusively of two-arm trials. However, approximately 85\% of network meta-analyses in the literature contain multi-arm trials \cite{Niko:2014}. We therefore test if our findings generalise to networks including multi-arm trials. We focus  on complete-loop networks (Figure~\ref{fig:networks}(c)) as this allows us to introduce three-arm and four-arm trials without changing the overall shape of the network (a full loop remains a full loop if further trials are added to it). The networks simulated in this section are designed specifically to cover a wide range of degree irregularities. We note that including more multi-arm trials in a network will, in general,  reduce its irregularity.

For a given value of the degree irregularity, $h^2/\hat{k}^2$, we generated synthetic trial data on complete-loop networks with different combinations of two-arm, three-arm and four-arm trials. We focus on the case of equally effective treatments, and report the outcome at network-level. Treatment-specific outcomes are provided in the Supplementary Material \cite{Davies:Supp} (Figures S25 to S47). In all cases, the relationship between bias of rank probability and the number of studies per treatment follows the same pattern as in Figure \ref{fig:RankProb_multi}.

\begin{figure}
    \centering
    \includegraphics[width=0.94\linewidth]{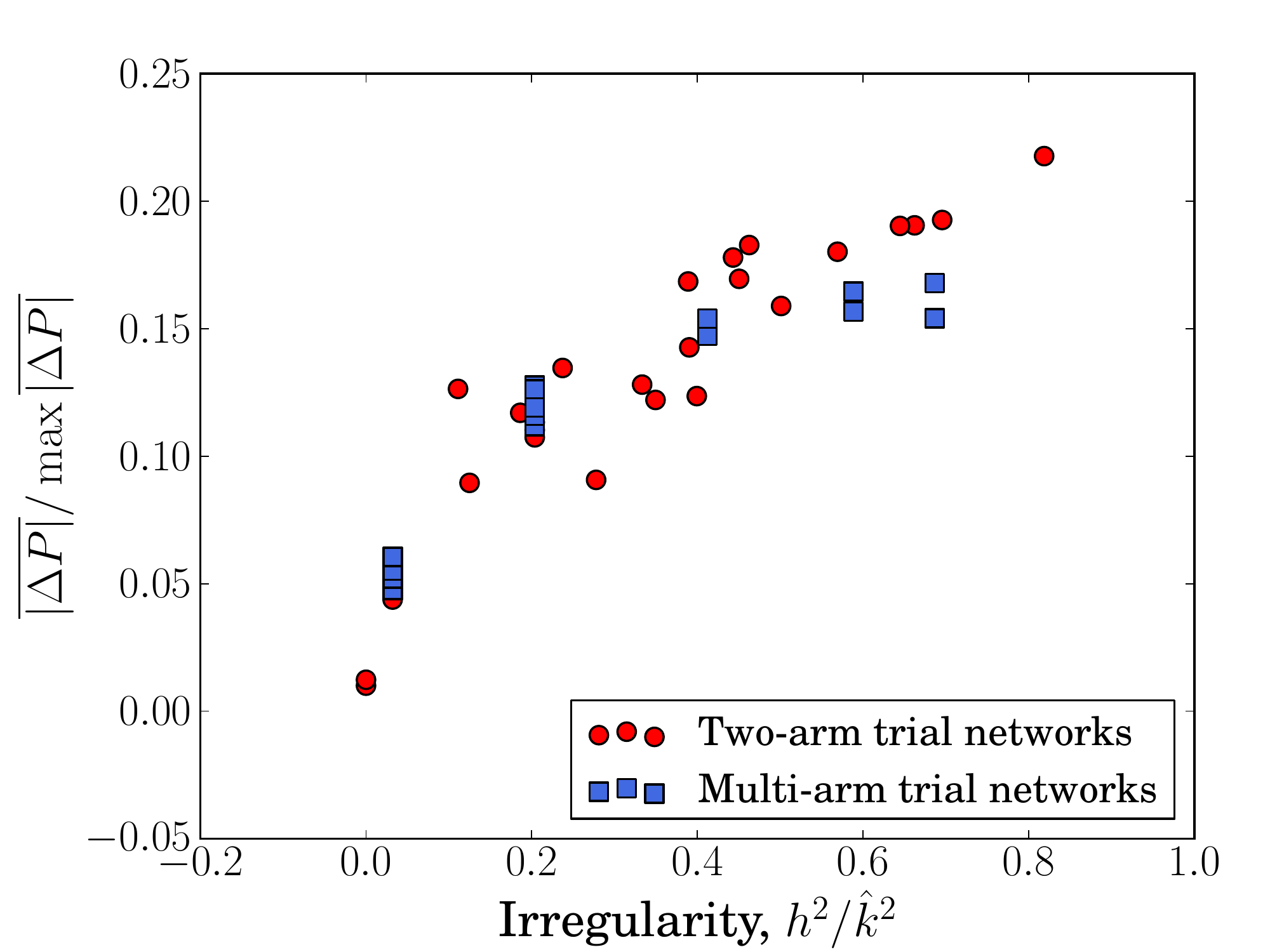}
    \caption{~  The effect of degree irregularity on a network's total rank probability bias. Data from networks with multi-arm trials is shown as blue squares. }
    \label{fig:h_RankProb_MultiArm}
    \end{figure}

    \begin{figure}
    \includegraphics[width=0.94\linewidth]{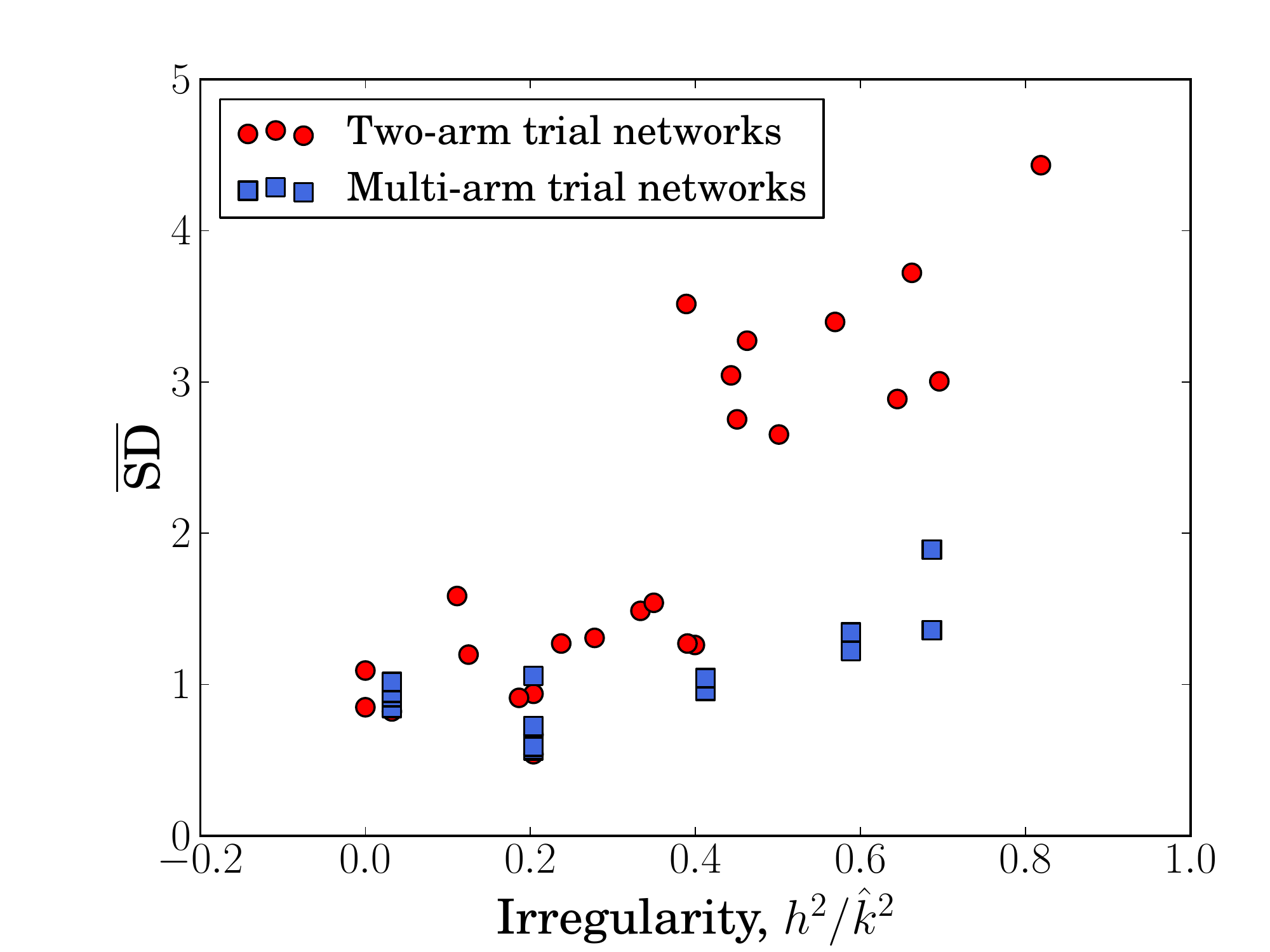}
    \caption{~  The effect of degree irregularity on a network's total standard deviation of treatment effect estimates. Data from networks with multi-arm trials is shown as blue squares. }
    \label{fig:h_SD_multi}
    \end{figure}

    \begin{figure}
    \includegraphics[width=0.94\linewidth]{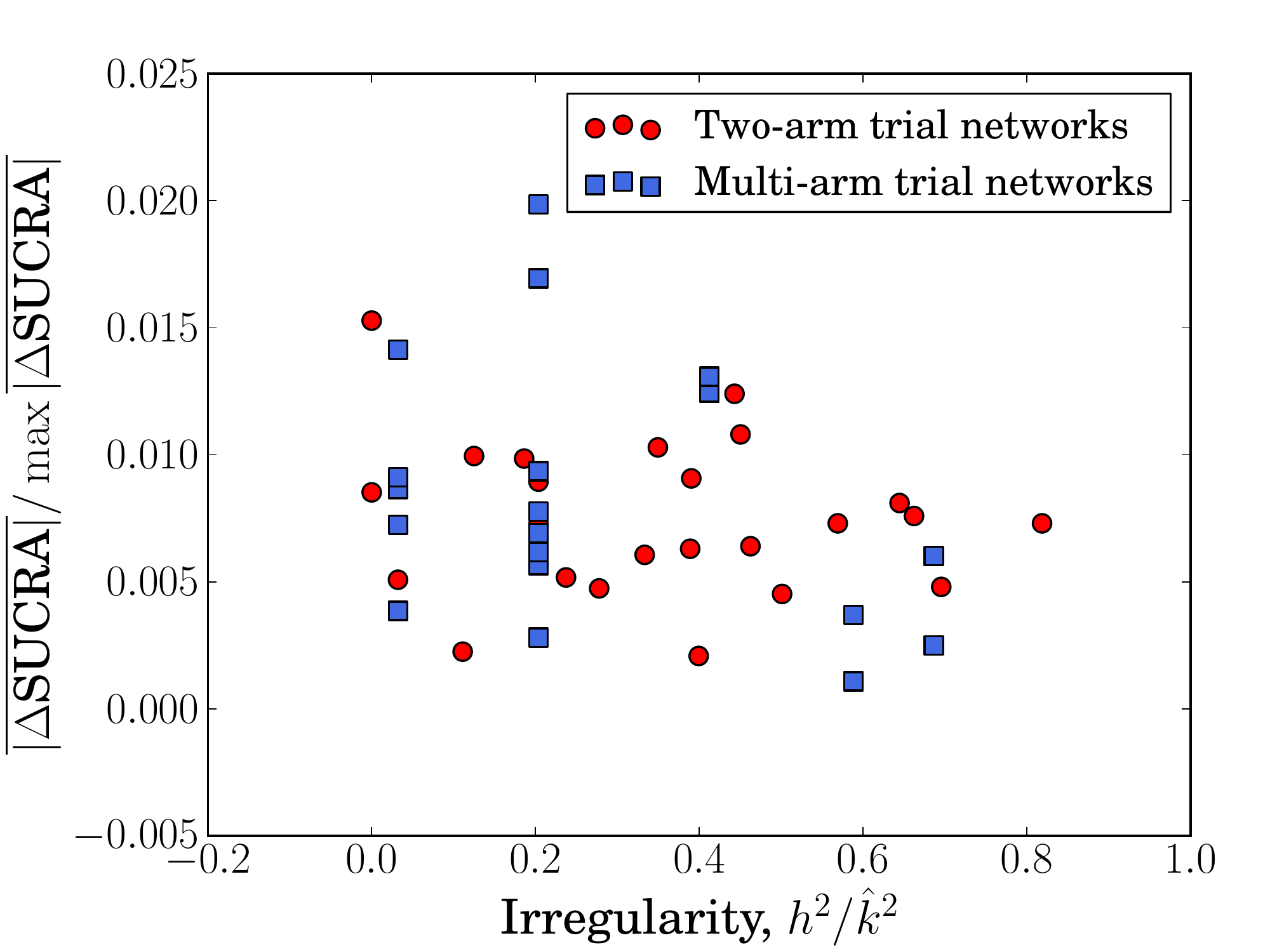}
    \caption{~  The effect of degree irregularity on a network's total bias on SUCRA values. Data from networks with multi-arm trials is shown as blue squares. }
    \label{fig:h_SUCRA_multi}
\end{figure}

\begin{figure*}%
    \centering
    \includegraphics[width=0.44\linewidth]{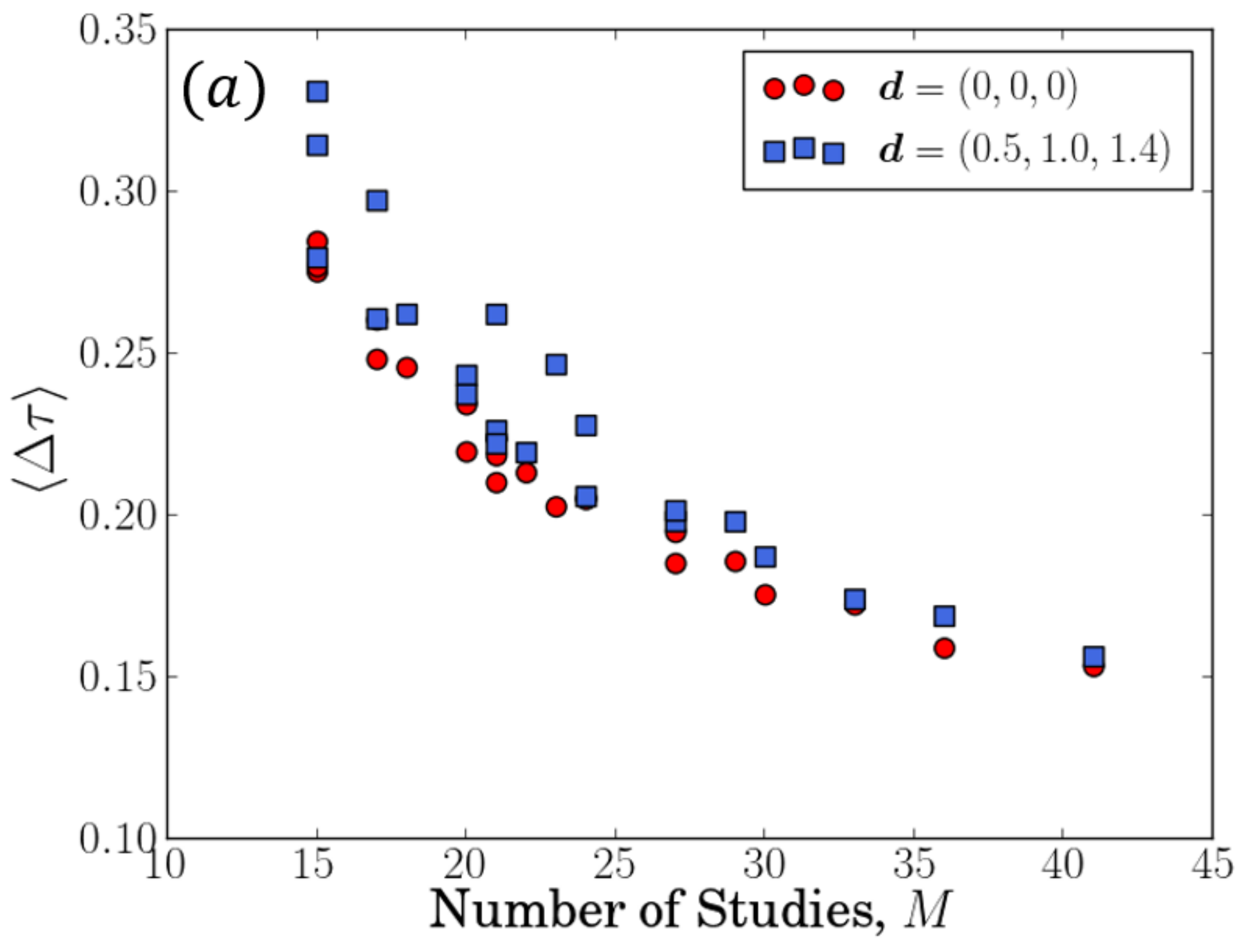} 
    \qquad
    \includegraphics[width=0.44\linewidth]{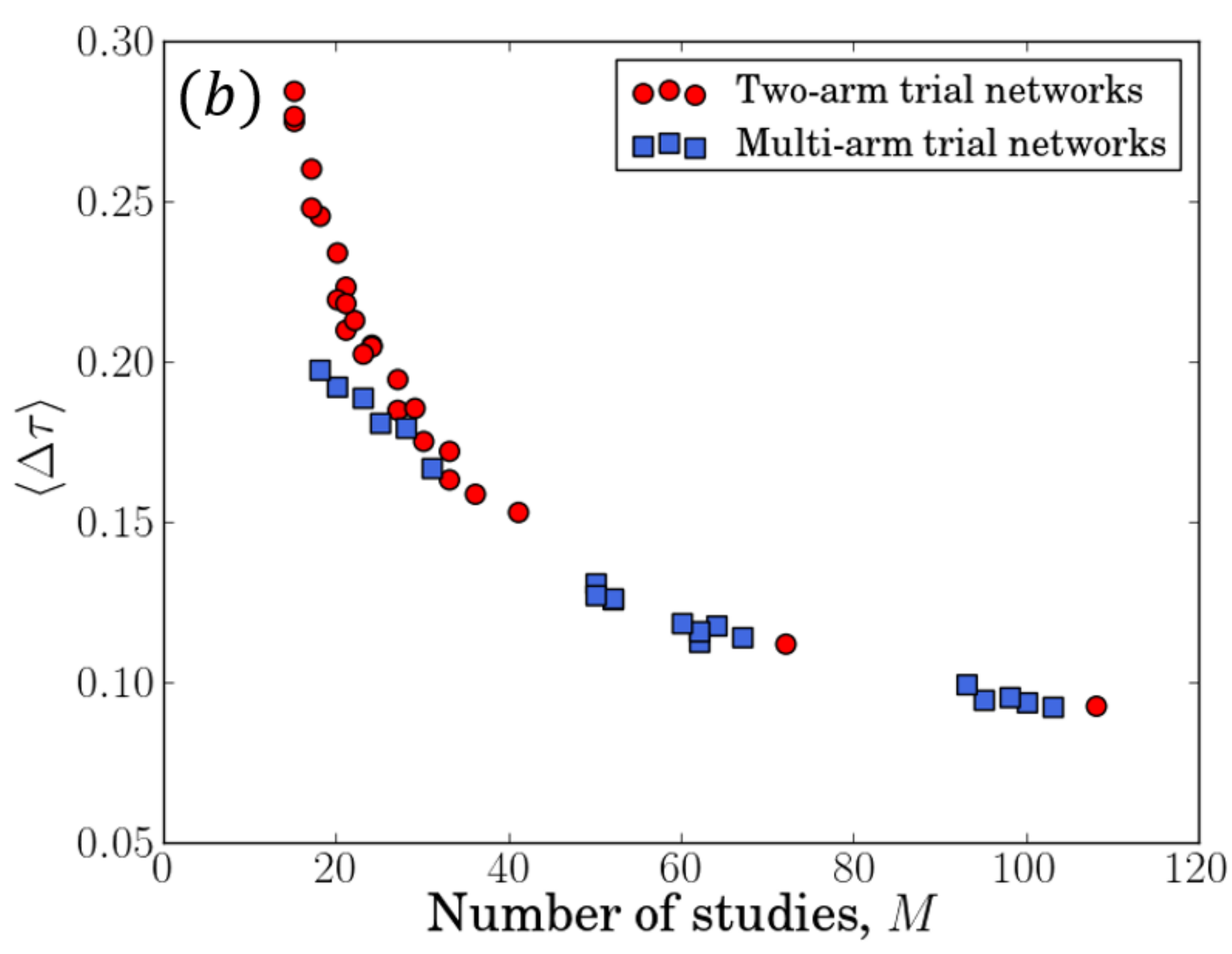} 
    \caption{~  The effect of the total number of studies in a network on the accuracy of the estimate for the heterogeneity parameter $\tau$. Panel (a) is for networks made up exclusively of two-arm trials and compares networks with equally effective and non-equally effective treatments. Panel (b) includes networks with $\boldsymbol{d}=(0,0,0)$ only and networks with multi-arm trials are shown as blue squares. } \label{fig:tau}
\end{figure*}

We show $\overline{|\Delta P|}$, $\overline{\text{SD}}$ and $\overline{|\Delta \text{SUCRA}|}$ as a function of network irregularity in Figures \ref{fig:h_RankProb_MultiArm} to \ref{fig:h_SUCRA_multi} respectively. The data from networks involving multi-arm trials is indicated by blue squares; we include the data for networks of two-arm trials (red circles) to allow comparison. As for the case of two-arm trials, the total magnitude of the bias of rank probabilities and the total standard deviation of treatment effects increase with  $h^2/\hat{k}^2$, while the bias on SUCRA is largely unaffected by network irregularity. When networks are sufficiently irregular, the presence of multi-arm trials appears to reduce $\overline{\text{SD}}$  with respect to networks consisting only of two-arm trials (Figure \ref{fig:h_SD_multi}).

These results show that our findings concerning both within-network and between-network comparisons can be generalised to networks containing multi-arm trials.

\subsection{Data-generating models}
All data so far was produced using the data-generating model `Normal' (see Sec.~\ref{DGM}). We also carried out a similar analysis using data from the `Euclidean' and `Uniform' methods.  The only difference we observe is in the magnitude of the standard deviation of treatment effects. While the relationship between network irregularity, $h^2/\hat{k}^2$, and total standard deviation, $\overline{\text{SD}}$, was not affected by the choice of DGM, the values of $\overline{\text{SD}}$ were lowest for the `Euclidean' method and highest for the `Uniform' method. This is not surprising as the `Euclidean' method restricts the range of absolute treatment effects that can be sampled and thus reduces variation in the event rates. The `Uniform' method is the least restrictive in this sense. All other results were consistent between the three DGMs (see Figures S48 to S50 in the Supplementary Material \cite{Davies:Supp}). This demonstrates that the effects we observe are due to the network geometry, and are not specific to any data-generating model. 

\subsection[Bias of the heterogeneity parameter]{Bias of the heterogeneity parameter, $\tau$}

The data in Figure~\ref{fig:tau} shows that bias of the heterogeneity parameter, $\tau$, decreases with the total number of studies in the network. This is the case irrespective of whether the treatments have uniform or varying effects ($\boldsymbol{d}=\boldsymbol{0}$ or $\boldsymbol{d}\neq \boldsymbol{0}$), and for networks with two-arm and multi-arm trials. The bias of $\tau$ is not affected by the network's irregularity $h^2/\hat{k}^2$ (see Figure S51 in the Supplementary Material \cite{Davies:Supp}). Therefore adding any trial to the network improves the accuracy of the estimate of $\tau$. 

The true value of $\tau$ in Figure~\ref{fig:tau} is $0.1$; we note that $\tau$ is considerably overestimated in all cases we tested. To understand this, it is useful to recall that $\tau$ characterises the variation of the relative effects between any two treatments across trials in the random-effects model. Additional randomness originates from the sampling of event numbers in each trial arm. This is the case both in real-world trial data and in simulation studies (in the latter the sampling is from the binomial distributions for the respective trial arms). Some of this sampling noise may be attributed to between-trial variability by the NMA method, leading to an overestimation of $\tau$.

\section{Summary and Discussion}
\label{discuss}
\subsection{Variation of treatment effect uncertainty is associated with biased rank probabilities}
\label{SD_Rank}
We have carried out simulation studies of network meta-analysis in random-effects models. These simulations reveal that disparity in the number of studies different treatments are involved in can lead to variation between the standard deviations of effect estimates. This in turn appears to generate a systematic bias in estimated rank probabilities. In line with previous simulations of NMA for fixed-effects models \cite{Kibret:2014}, the probability of a treatment being ranked best is overestimated for treatments included in the fewest number of studies, and underestimated for treatments which are part of a large number of studies.  In addition, our study of networks with four treatments found the same trend for the probability of being ranked last. The probability of being ranked second and third best is subject to a bias in the opposite direction. These trends correspond to an increased standard deviation of treatment effect estimates for treatments compared in a smaller number of studies.

\begin{figure}
    \centering
    \includegraphics[width=1.0\linewidth]{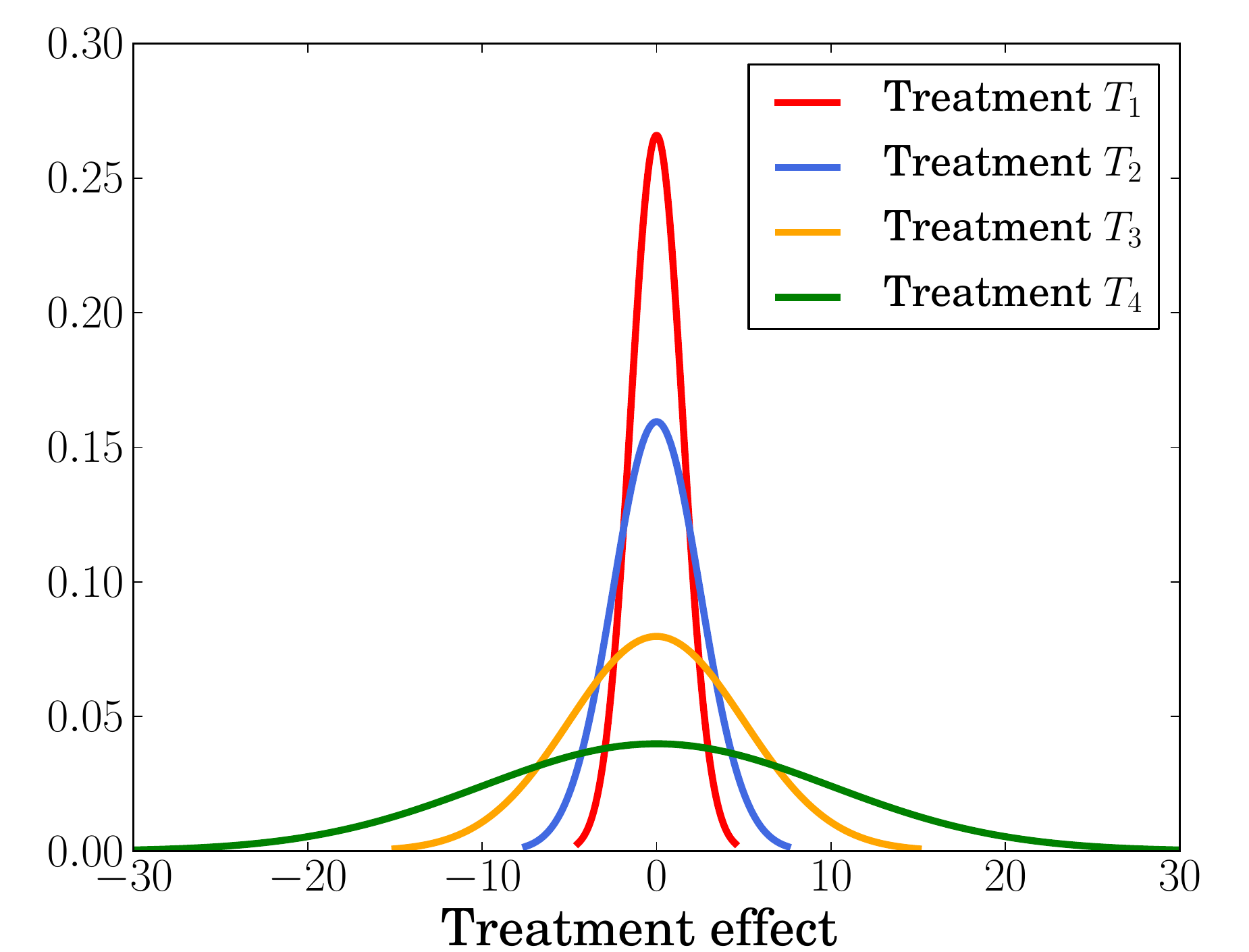}
    \caption{~  Illustrative example of posterior distributions of treatment effect estimates for four treatments in a network meta-analysis. The posterior distributions have the same mean value but varying standard deviation. Treatment $T_1$ has the most narrow distribution, followed by $T_2$, $T_3$ and $T_4$ which has the widest distribution. }
    \label{fig:Gaussians}
\end{figure}

A general connection between standard deviation of effect estimates and bias of rank probabilities has previously been recognised in  R\"{u}cker et al (2015) \cite{Rucker2015}. Our work establishes further details of the mechanics leading to biased rank probabilities. We illustrate this in Figure \ref{fig:Gaussians}, where we show a fictitious example of posterior distributions for the effectiveness of four different treatments. The four distributions have equal mean values, but varying standard deviations. The distribution of treatment $T_4$, which has the largest standard deviation, has higher density than the other treatments at very large and very small values of the treatment effect. This means that although the most probable value of the effect of treatment $T_4$ is the same as for the other treatments, $T_4$ is more likely than the other treatments to have an effect that is the largest or the smallest. Therefore treatment $T_4$ has the highest probability of being ranked best and the highest probability of being ranked worst. Conversely, treatment $T_1$ has the lowest standard deviation. Therefore, it is less likely to have extreme values of treatment effect and thus has a higher probability of being ranked second or third. R\"{u}cker et al (2015) \cite{Rucker2015} used a similar explanation to demonstrate that the probability that one treatment is better than another can be misleading when the posterior distributions of their effects have considerable overlap. 

This stylised example demonstrates that biased rank probabilities can result if the uncertainty on some treatment effects is larger than on others. This effect is also to be expected when the distributions of treatment effects have different means, provided the differences in these means are small compared to their standard deviations. 

Our analysis shows that the posterior distributions of treatment effect estimates are the most narrow for treatments included in the most studies and widest for those that have been studied the least. As a consequence, biases of rank probabilities may arise if different treatments are involved in disparate numbers of studies, i.e., for large irregularity of the network.

The simulations presented in this paper are an explicit demonstration of biases that can occur in the comparison of multiple treatments. We have also suggested how they might originate from the structure of the network of treatments and trials. Understanding the origins of bias, we think, is vital for interpreting rank probabilities in network meta-analyses, and contributes to our understanding of the NMA method and its limitations.

\subsection{Planning future studies to reduce the irregularity of the network}
\label{Plan_future}

Planning future clinical trials based on existing evidence from network meta-analysis can reduce the resources and number of participants required to obtain results of a given precision \cite{Salanti:2018, Nikola:2015, Ioannidis:2014}. While the design of future trials based on pairwise meta-analysis has received significant attention \cite{Sutton:2007, Sutton:2009, Roloff:2013}, methods using the outcome of network meta-analysis are less developed. Current approaches in this area \cite{Nikola:2014, Nikola:2015} are computationally intensive and become increasingly laborious as the network becomes more complex. Our results show that the degree irregularity of a network, $h^2/\hat{k}^2$, can provide guidance on the choice of future trials without the need for extensive simulations. The degree of a treatment in the graph is the number of trials it is involved in, and the irregularity of a network describes how this degree varies across treatments. This is easily obtained from the network.

\begin{figure}
    \centering
    \includegraphics[width=1.0\linewidth]{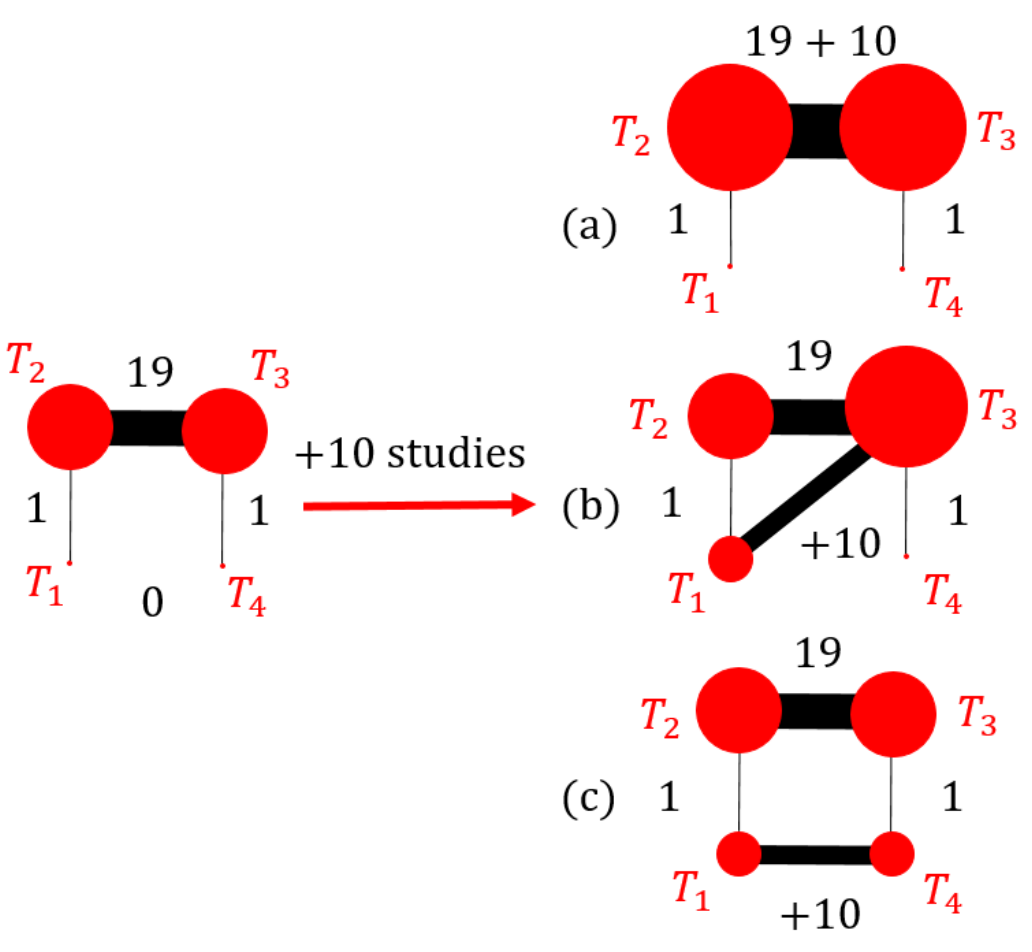}
    \caption{~  Example of three different geometries that can be created by adding a fixed number of studies to an existing network. }
    \label{fig:plan}
\end{figure}

Irregularity is a better indicator of the quality of a network meta-analysis than the total number of studies in the network. As we have shown, networks with a more homogeneous distribution of studies between treatments have more precise treatment effect estimates and smaller bias of rank probabilities. 

Degree irregularity is therefore a useful metric for working out which comparisons could be made in future studies to improve the quality of an existing NMA. For example, consider a network of four treatments with $\boldsymbol{K}=(1,0,0,19,0,1)$ and $h^2/\hat{k}^2 = 0.82$ as shown on the left in Figure \ref{fig:plan}. Now imagine resources are available to add ten new two-arm studies to this network. If (a) we add all ten studies to the most connected comparison ($T_2-T_3$) then we obtain $\boldsymbol{K}=(1,0,0,29,0,1)$, and the network's irregularity increases to $0.88$. We may be more inclined to populate a comparison that currently has no direct evidence such as $T_1-T_3$ [(b) in Figure \ref{fig:plan}] or $T_1-T_4$ [(c)]. The former leads to $\boldsymbol{K}=(1,10,0,19,0,1)$ and reduces $h^2/\hat{k}^2$ to $0.48$, while the latter has $\boldsymbol{K}=(1,0,10,19,0,1)$ and reduces $h^2/\hat{k}^2$ to $0.08$. These three possible `future' networks are shown on the right-hand side in Figure \ref{fig:plan}.

By simulating the original network and the three `future' networks whilst keeping all other network characteristics constant, we compare how adding the extra ten studies affects the quality of the results. Table \ref{egdata} summarises the total standard deviation and total rank probability bias of these four networks. 

\begin{table}[h!]
\caption{~  Degree heterogeneity and quality of NMA outcome for the networks in Figure \ref{fig:plan}.}
\centering
\begin{tabular}{l c c c c}
\toprule
Network   & $M$ & $h^2/\hat{k}^2$ & $\overline{\text{SD}}$ & $\overline{|\Delta P|}$\\
\midrule
Original: & 21 & 0.82 & 4.44 & 1.74 \\
(a): & 31 & 0.88 & 4.36 & 1.73 \\
(b): & 31 & 0.48 & 3.24 & 1.47 \\
(c): & 31 & 0.08 & 1.68 & 0.16 \\
\bottomrule
\end{tabular}
\label{egdata}
\end{table}

For network (a) the quality of the NMA is approximately the same as for the original network whereas (b) and (c) show a considerable reduction in $\overline{\text{SD}}$ and $\overline{|\Delta P|}$. The improvement in both quantities for network (c) is markedly greater than in network (b) even though in both cases the ten new studies were added to a comparison with no existing direct evidence. 

This example demonstrates that equality in the number of studies per treatment is more important than equality in the number of studies per comparison. Choosing future studies that reduce degree irregularity may therefore help to improve the precision of treatment effect estimates and the accuracy of rank probabilities.

\subsection*{Acknowledgements}

AD acknowledges funding by the Engineering and Physical Sciences Research Council (EPSRC UK), grant number EP/R513131/1. TG is grateful for partial financial support by the Maria de Maeztu Program for Units of Excellence in R\&D (MDM-2017-0711).

\end{document}